\documentstyle[prd,aps,preprint,tighten,psfig]{revtex}

\begin{document}
\draft

\title{Charmonium in the instantaneous approximation}

\author{Johan Linde\footnote{E-mail: jl@theophys.kth.se} and H{\aa}kan 
Snellman\footnote{E-mail: snell@theophys.kth.se}} \address{Department 
of Theoretical Physics\\
Royal Institute of Technology\\
S-100 44 STOCKHOLM\\
SWEDEN}

\maketitle

\begin{abstract}
The charmonium system is studied in a Salpeter model with a vector 
plus scalar potential.  We use a kinematical formalism based on the 
one developed by Suttorp\cite{Suttorp}, and present general eigenvalue 
equations and expressions for decay observables in an onium system for 
such a potential both in the Feynman and Coulomb gauges.  Special 
attention is paid to the problem with renormalization of the lepton 
pair decays, and we argue that they must be defined relative to one of 
the experimental decay widths because renormalization of the vertex 
function is not possible.  The parameters of the model are determined 
by a fit to the mass spectrum and the lepton pair decay rates.  Two 
gamma decays and E1 and M1 transitions are then calculated and found 
to be well accounted for.  No significant differences in the results 
in Feynman or Coulomb gauge are found.  A comparison is made, 
regarding the electromagnetic transitions, between the full and 
reduced Salpeter equation.  A large difference is found showing the 
importance of using the full Salpeter equation.
\end{abstract}


\narrowtext

\section{Introduction}
The Bethe-Salpeter formalism has recently been used in various ways in 
more careful investigations of quarkonium physics.  For example 
Munczek and Jain\cite{munczek} have used the full Bethe-Salpeter 
equation together with the Schwinger-Dyson equation to study the mass 
spectrum and lepton pair decays of light and medium-heavy mesons, 
Resag {\it et 
al.\/}\cite{ResagochMunz1,ResagochMunz2,ResagochMunz3,ResagochMunz4}
have written a series of papers about the Salpeter equation where they 
have studied mass spectra and various decay observables for both 
light and heavy mesons, Kugo {\it et al.\/}\cite{kugo} have calculated 
the Isgur-Wise function and Harada and Yoshida have calculated the QCD 
$S$ parameter\cite{harada1} and developed a method to solve the full 
Bethe-Salpeter equation\cite{harada2}.  The stability of the Salpeter 
equation for various Lorentz structures of the confining potential 
has been studied by Olsson {\it et al.\/}\cite{olsson_stabil} and 
Parramore {\it et al.\/}\cite{parramore1,parramore2}.  Olsson {\it et 
al.\/} have also studied the validity of the reduced Salpeter 
equation\cite{olsson_2}.

In this paper we undertake an investigation of the radiative decays of 
the lowest charmonium levels including both M1 and E1 transitions in 
our study as well as the two-gamma decays of the corresponding 
pseudoscalar, scalar and tensor states.  

Like many other authors\cite{ResagochMunz1,murota,Ito_E1} we use the 
instantaneous approximation of the Bethe-Salpeter equation -- the 
Salpeter equation -- which has a well defined physical interpretation.  
Our kinematical analysis uses a formalism based on the one developed 
by Suttorp\cite{Suttorp}.  (Another formalism in a similar spirit has 
been developed by Laga\"e\cite{Lagae}.)  This leads to a set of 
coupled equations for the amplitudes which are solved numerically.  In 
this paper we present a general set of eigenvalue equations and decay 
amplitudes for lepton pair decays, two gamma decays and E1 and M1 
transitions, which can be used in any model with a vector plus scalar 
potential using the formalism of Suttorp.

Our potential consists of one short-range vector part describing the 
one-gluon-exchange interaction and one scalar long-range confining 
part.  Investigations of the stability of the Salpeter equation for 
confining potentials with various Lorentz structures have shown that 
for light quarks there are no stable solutions for a scalar 
confinement\cite{olsson_stabil,parramore1,parramore2}. This problem 
does not appear for heavy quarks like the charm quark.  Furthermore a 
Lorentz time component vector confinement, which has been shown to be 
the most stable one, gives a contribution to the spin-orbit coupling 
with the wrong sign.  We have therefore chosen a scalar confinement 
potential.

In quantum mechanics the spectrum itself does not uniquely determine 
the potential.  We also need as input some other information 
concerning the localization of the system, such as the value of the 
wave function close to the origin.  As input we therefore take the 
energy levels of the charmonium spectrum as well as the decay rates of 
the vector $1^{--}$~states into lepton pairs.

The instantaneous approximation leads, strictly speaking, to some 
problems with renormalization, since the Ward-Takahashi identities are 
then not relevant for the renormalization of the vertex function in 
the lepton pair decay.  Furthermore the kernel is confining and 
contains a dimensionful coupling constant for the confining part.  
This leads in the final analysis to a renormalization procedure where 
we use the decay width for $J/\Psi\to e^{+}e^{-}$ to define the 
renormalization constant.  Thus, we calculate the ratios of the 
decay rates of the vector states into lepton pairs, rather than the 
rates themselves.

We adopt the formalism of Mandelstam\cite{Mandelstam} to calculate the 
radiative decay amplitudes.  The gauge invariance of this method in 
the instantaneous approximation is not obvious, but in Sec.  
\ref{m1e1decay} we show that the decay amplitudes are indeed gauge 
invariant.

The paper on heavy quarkonia by Resag and M\"unz\cite{ResagochMunz3} 
overlaps to some extent with our paper.  They have also used a Salpeter 
model to calculate mass spectra and decay observables for charmonium.  
However, they have used a somewhat different potential.  In 
particular, they have introduced a cut-off in the potential at the 
origin in order to renormalize the lepton pair decay widths.  In this 
paper we discuss the difficulties of renormalization and conclude that 
an experimental decay width has to be used to properly define the 
renormalization. A cut-off also necessitates some extra 
parameters in the model, and they obviously affect the 
results. In particular, the lepton pair decay widths depend on the 
wave function's value at the origin, and therefore one can obtain any 
value for those by varying the cut-off parameters.

In addition, we present general equations in a formalism different 
from\cite{ResagochMunz3} and investigate the importance of non-dominant 
terms in the wave functions.

The outline of our paper is as follows.  In Section~\ref{salpeq} we 
briefly discuss the Bethe-Salpeter equation to define our variables.  
In Section~\ref{potential} we describe the potential used in the 
model.  We use a short-range Coulomb potential plus a long-range 
confining linear potential.  In Section~\ref{leptondecay} we give the 
expression for lepton pair decay and discuss the problem with 
renormalization.  In Section~\ref{ggdecay} we give the expressions 
for the $\gamma\gamma$ decays and in Section~\ref{m1e1decay} we give 
the expressions for the M1 and E1 transitions.  There we also discuss 
the gauge invariance in the instantaneous approximation.  In 
Section~\ref{resultat} we present the numerical results and finally in 
Section~\ref{tillsist} we give a summary of our results and 
conclusions.

\section{The Salpeter equation} 
\label{salpeq}
The wave function $\chi(q)$ for bound states of a fermion-antifermion 
pair satisfies the spinor Bethe-Salpeter equation
\begin{equation}
(\rlap{/}{q}+\frac{\rlap{/}{P}}{2}-m)\chi(q)
(\rlap{/}{q}-\frac{\rlap{/}{P}}{2}-m)= 
i\sum_{i=1}^5\int\frac{d^4p}{(2\pi)^4}K^i(p,q,P)\Gamma^i\chi(p)\Gamma^i.
\label{bs-eq}
\end{equation}
Here, $P$ is the momentum of the bound state and $q$ is the relative 
momentum of the constituent fermions.  The interaction kernel is 
characterized by the matrices $\Gamma^i$, where $\Gamma^S=1$, 
$\Gamma^V=\gamma^\mu$, $\Gamma^T=\sigma^{\mu\nu}$, 
$\Gamma^A=\gamma^\mu\gamma^5$, $\Gamma^P=\gamma^5$.

We now make the instantaneous approximation, so that the kernels 
$K^i=K^i({\bf p}-{\bf q})$ in the rest frame.  This reduces the 
Bethe-Salpeter equation to the Salpeter equation.  It may seem like 
Lorentz invariance is lost by making this approximation.  However, 
although manifest Lorentz invariance is lost, the Salpeter equation 
can be formulated covariantly by letting $K^i$ depend on 
$q_\perp=q-\frac{q\cdot P}{M^2}P$ and $p_\perp=p-\frac{p\cdot 
P}{M^2}P$ in a general frame.  In the rest frame $q_\perp$ reduces to 
$(0,{\bf q})$. For details, see for example Chang and Chen\cite{kineser}.

It is most convenient to calculate the mass spectrum and wave 
functions in the rest frame of the bound states, and afterwards 
Lorentz transform the wave functions when needed.  We will therefore
perform the calculations in the rest frame
for the remainder of the paper.

The propagators can be written
\begin{eqnarray}
\frac{1}{\rlap{/}{q}-m}=&&\left(\frac{\Lambda_{+}^{+}({\bf q})}{(q^0-E({\bf 
q})+i\epsilon)}+ \frac{\Lambda_{-}^{+}({\bf q})}{(q^0+E({\bf 
q})-i\epsilon)}\right)\gamma^0\nonumber\\
=&&\gamma^0\left(\frac{\Lambda_{+}^{-}({\bf q})}{(q^0-E({\bf 
q})+i\epsilon)}+ \frac{\Lambda_{-}^{-}({\bf q})}{(q^0+E({\bf 
q})-i\epsilon)}\right),\label{propagator}
\end{eqnarray}
where
\begin{equation}
\Lambda_\mu^\pm({\bf q})=\frac{1}{2E({\bf q})}(E({\bf 
q})+\mu(\pm\bbox{\alpha}\cdot{\bf q}+m\gamma^0)) \qquad \mu=\pm
\end{equation}
are the projection operators for the positive and negative energy 
states respectively.

With the help of these projection operators we can write the Salpeter 
equation as a set of four coupled equations,
\begin{eqnarray}
\Lambda_\mu^+({\bf q})\chi(q)\Lambda_\nu^-({\bf q})=&& 
\frac{i}{(M/2+q^0-\mu E+\mu i\epsilon)(-M/2+q^0-\nu E+\nu 
i\epsilon)}\times \nonumber\\
&&\quad\sum_i\int\frac{d^4p}{(2\pi)^4}K^i({\bf p}-{\bf q}) 
\Lambda_\mu^+({\bf q})\gamma^0\Gamma^i\chi(p)\Gamma^i\gamma^0 
\Lambda_\nu^-({\bf q}). \label{fyrachiekv}
\end{eqnarray}
The right hand side can be integrated over $q^0$ by a simple contour 
integration.  Introducing the reduced wave function
\begin{equation}
\Psi({\bf q})=\int\frac{dq^0}{2\pi}\chi(q) \label{redwave}
\end{equation}
we find that
\begin{eqnarray}
\Lambda_{+}^{+}({\bf q})\Psi({\bf q})\Lambda_{-}^{-}({\bf q})=&& 
-\frac{1}{(M-2E)} \sum_i\int\frac{d{\bf p}}{(2\pi)^{3}} K^i({\bf 
p}-{\bf q}) \Lambda_{+}^{+}({\bf q})\gamma^0\Gamma^i\Psi({\bf 
p})\Gamma^i\gamma^0 \Lambda_{-}^{-}({\bf q}), \label{psieq1}\\
\Lambda_{-}^{+}({\bf q})\Psi({\bf q})\Lambda_{+}^{-}({\bf q})=&& 
\frac{1}{(M+2E)} \sum_i\int\frac{d{\bf p}}{(2\pi)^{3}} K^i({\bf 
p}-{\bf q}) \Lambda_{-}^{+}({\bf q})\gamma^0\Gamma^i\Psi({\bf 
p})\Gamma^i\gamma^0 \Lambda_{+}^{-}({\bf q}),\label{litet_psi}\\
\Lambda_{+}^{+}({\bf q})\Psi({\bf q})\Lambda_{+}^{-}({\bf q})=&&0 ,
\label{psieq3}\\
\Lambda_{-}^{+}({\bf q})\Psi({\bf q})\Lambda_{-}^{-}({\bf q})=&&0 .
\label{psieq4}
\end{eqnarray}
Combining these equations with (\ref{fyrachiekv}), we can write the 
wave function as
\begin{eqnarray}
\lefteqn{\chi(q)={}}\nonumber\\
=&&-\frac{i(M-2E)}{(M/2+q^0-E+i\epsilon)(-M/2+q^0+E-i\epsilon)} 
\Lambda_{+}^{+}({\bf q})\Psi({\bf q})\Lambda_{-}^{-}({\bf 
q})\nonumber\\
&&{}+\frac{i(M+2E)}{(M/2+q^0+E-i\epsilon)(-M/2+q^0-E+i\epsilon)} 
\Lambda_{-}^{+}({\bf q})\Psi({\bf q})\Lambda_{+}^{-}({\bf 
q})\nonumber\\
&&{}+\frac{i}{(M/2+q^0-E+i\epsilon)(-M/2+q^0-E+i\epsilon)} 
\int\frac{d{\bf p}}{(2\pi)^{3}} 
K^i\Lambda_{+}^{+}({\bf q})\gamma^0\Gamma^i\Psi({\bf 
p})\Gamma^i\gamma^0\Lambda_{+}^{-}({\bf q})\nonumber\\
&&{}+\frac{i}{(M/2+q^0+E-i\epsilon)(-M/2+q^0+E-i\epsilon)} 
\int\frac{d{\bf p}}{(2\pi)^{3}} 
K^i\Lambda_{-}^{+}({\bf q})\gamma^0\Gamma^i\Psi({\bf 
p})\Gamma^i\gamma^0\Lambda_{-}^{-}({\bf q}). \label{waveeq}
\end{eqnarray}
Thus, finding the wave function $\chi(q)$ in this formalism reduces to 
finding $\Psi({\bf q})$.

Combining the equations (\ref{psieq1})-(\ref{psieq4}) we get an equation 
for $\Psi({\bf q})$,
\begin{eqnarray}
\lefteqn{(\frac{M}{2}-\bbox{\alpha}\cdot{\bf q}-m\gamma^0)\Psi({\bf 
q})+\Psi({\bf q}) (\frac{M}{2}-\bbox{\alpha}\cdot{\bf 
q}+m\gamma^0)}\nonumber\\
&&=-\sum_i\int\frac{d{\bf p}}{(2\pi)^{3}} K^i({\bf p}-{\bf 
q})\bigl(\Lambda_{+}^{+}({\bf q})\gamma^0\Gamma^i\Psi({\bf p}) 
\Gamma^i\gamma^0\Lambda_{-}^{-}({\bf q})-\Lambda_{-}^{+}({\bf 
q})\gamma^0\Gamma^i\Psi({\bf p})\Gamma^i\gamma^0 \Lambda_{+}^{-}({\bf 
q})\bigr) .\label{psiequation}
\end{eqnarray}

For each state with a certain angular momentum, parity, and charge 
parity, we expand the function $\Psi({\bf q})$ in the Dirac algebra by 
imposing the correct transformation properties.  We also demand that 
$\Psi({\bf q})$ fulfills the conditions (\ref{psieq3})-(\ref{psieq4}).  
For example the expansion of the $0^{-+}$~state is
\begin{equation}
\Psi({\bf q})=\Psi_P(q)\gamma^5+\Psi_A(q)\left(\gamma^0\gamma^5+ 
\gamma^5\frac{\bbox{\alpha}\cdot{\bf q}}{m}\right).
\end{equation}
The scalar functions $\Psi_A$ and $\Psi_P$ depend only on $q=|{\bf 
q}|$.  Similar expansions for other states are given in 
Appendix~\ref{waveexp}.

Integral equations for the scalar functions $\Psi_i$ can be obtained 
from (\ref{psiequation}) by expanding both sides in the Dirac algebra 
and comparing the coefficients.  This is discussed further in 
Appendix~\ref{eigenequ}, where all integral equations are given.

\section{The potential} 
\label{potential}
We have chosen a potential which consists of one short-range part 
describing the one-gluon-exchange interaction and one long-range 
confining part. The calculations have been done both in the Feynman 
gauge and in the Coulomb gauge, in an attempt to study the gauge 
dependence in the instantaneous approximation. The respective 
potentials are
\begin{eqnarray}
K_{\text{Feyn}}(r)=&&K^V(r)\gamma_\mu^{(1)}\gamma^{\mu(2)} 
+K^S(r)+U,\\
K_{\text{Col}}(r)=&&K^V(r)\left(\gamma^{0(1)}\gamma^{0(2)}-\frac{1}{2} 
\bbox{\gamma}^{(1)}\bbox{\gamma}^{(2)}+
\frac{1}{2}(\bbox{\gamma}^{(1)}\cdot{\bf\hat x})
(\bbox{\gamma}^{(2)}\cdot{\bf\hat x})\right) 
+K^S(r)+U,
\end{eqnarray}
where $U$ is a constant and
\begin{eqnarray}
	K^{V}(r) & = & \frac{\bar\alpha_s}{r},  \\
	K^{S}(r) & = & \lambda r .
\end{eqnarray}
In momentum space the vector part in the Coulomb gauge becomes
\begin{equation}
	\frac{4\pi\bar\alpha_{s}}{({\bf p}-{\bf q})^{2}} \left( 
	\gamma_\mu^{(1)}\gamma^{\mu(2)} +(\bbox{\gamma}^{(1)}\cdot({\bf \hat 
	p}-{\bf \hat q}))(\bbox{\gamma}^{(2)}\cdot
	({\bf \hat 
	p}-{\bf \hat q}))\right) \label{coulomb_mom}
\end{equation}
The color factor $4/3$ is included in $\bar\alpha_s$, 
$\bar\alpha_s=4/3\alpha_s$.  We let the coupling be constant, since a 
running coupling would not make any significant difference for heavy 
mesons like charmonium.  

There are four parameters in the model: $\bar\alpha_s$, $\lambda$, $U$,
and the quark mass $m$.  These will be determined by a least square 
fit to the mass spectrum and the ratios of the lepton pair decays.

Olsson {\it et al.\/}\cite{olsson_stabil} have recently investigated 
the stability of the Salpeter equation for various Lorentz 
structures of the confining kernel.  They have come to the conclusion 
that for light and medium-heavy quarks numerically stable solutions 
exist only for a time component vector potential.  The same result has 
also been found  by Parramore and 
Piekarewicz\cite{parramore1,parramore2} from a similar analysis.

However, we have found no such problems for heavy quarks like in the 
charmonium system. This has also been 
confirmed in Ref.\cite{ResagochMunz3}.  Furthermore, 
a time component vector confinement gives a contribution to the 
spin-orbit coupling with the wrong sign, which speaks in favor of a 
scalar confinement.

We want to be able to perform the angular integrations in the Salpeter 
equation analytically.  We therefore need the expansions of the 
potential in spherical harmonics in momentum space as
\begin{equation}
K^i({\bf p}-{\bf q})=\sum_{\ell m}F_\ell^i(p,q)Y^*_{\ell m}(\hat {\bf 
p})Y_{\ell m}(\hat {\bf q}), \quad i=V,S.
\end{equation}
The expansion of the Coulomb potential is
\begin{equation}
K^V({\bf p}-{\bf q})=\frac{4\pi\bar\alpha_s}{({\bf p}-{\bf q})^2}= 
\frac{8\pi^2\bar\alpha_s}{pq}\sum_{\ell 
m}Q_\ell\left(\frac{p^2+q^2}{2pq} \right)Y^*_{\ell m}(\hat {\bf 
p})Y_{\ell m}(\hat {\bf q}),
\end{equation}
where $Q_\ell$ are the Legendre functions of second kind.  Thus
\begin{equation}
F_\ell^V(p,q)=\frac{8\pi^2\bar\alpha_s}{pq}Q_\ell\left(\frac{p^2+q^2}{2pq} 
\right).
\end{equation}
In the Coulomb gauge potential (\ref{coulomb_mom}) there is another 
angular dependence in the second term.  However, we later choose to 
give this dependence in terms of the functions $F_{\ell}^{V}$.

The Fourier transform of the linear potential can only be defined in a 
distributional sense.  This causes no problems as in momentum space 
the value of the potential at each point is not important, but only its 
action on a wave function.  We define $K^S(r)$ as the limit
\begin{equation}
K^S(r)=\lim_{\epsilon\to0}\frac{\partial^2}{\partial\epsilon^2} 
\left(\frac{\lambda e^{-\epsilon r}}{r}\right).
\end{equation}
This gives
\begin{eqnarray}
K^S({\bf p}-{\bf 
q})=&&\lim_{\epsilon\to0}\frac{\partial^2}{\partial\epsilon^2}\left( 
\frac{4\pi\lambda}{({\bf p}-{\bf q})^2+\epsilon^2}\right) \nonumber \\
=&&\lim_{\epsilon\to0}\frac{8\pi^2\lambda}{pq}\sum_{\ell m} 
\frac{\partial^2}{\partial\epsilon^2}Q_\ell\left(\frac{p^2+q^2+\epsilon^2}{2pq} 
\right)Y^*_{\ell m}(\hat {\bf p})Y_{\ell m}(\hat {\bf q}).
\end{eqnarray}
It can be shown~\cite{Hersbach}, that in a distributional sense, the 
following relation holds:
\begin{eqnarray}
\lefteqn{\lim_{\epsilon\to0}\int_0^\infty 
dp\frac{\partial^2}{\partial\epsilon^2} 
Q_\ell\left(\frac{p^2+q^2+\epsilon^2}{2pq}\right)f(p)} \nonumber \\ 
&&=\text{PV}\int_0^\infty 
dp\left(\frac{1}{pq}Q'_\ell\left(\frac{p^2+q^2}{2pq} 
\right)f(p)+\frac{4q^2}{(p+q)^2(p-q)^2}f(q)\right),
\end{eqnarray}
where PV means the principal value.  Thus
\begin{eqnarray}
\lefteqn{\int_0^\infty\frac{dp}{(2\pi)^{3}} p^2 F_\ell^S(p,q)\Psi(p)}\nonumber \\ 
&&= \text{PV}\int_0^\infty \frac{dp}{(2\pi)^3}8\pi^2\lambda 
\left(\frac{1}{q^2}Q'_\ell\left(\frac{p^2+q^2}{2pq} 
\right)\Psi(p)+\frac{4q^2}{(p+q)^2(p-q)^2}\Psi(q)\right).
\end{eqnarray}

\section{Lepton pair decays} 
\label{leptondecay}
We now continue with a discussion of the decay observables.  A general 
prescription for calculating matrix elements of the electromagnetic 
current operator between bound states has 
been given by Mandelstam~\cite{Mandelstam}.  We begin with the lepton 
pair decays, which we will use together with the mass spectrum, in 
order to optimize the four parameters $\bar\alpha_s$, $\lambda$, $U$, 
and $m$ in the model.  The probability for a vector particle decaying 
into two light leptons is
\begin{equation}
\Gamma=\frac{4\pi\alpha^2 e_q^2}{M^3}|j^\mu j_\mu|^2,
\end{equation}
where $j^\mu$ is the vector quark current, $e_q$ is the charge of the 
quark, and $M$ is the mass of the vector particle.  To lowest order in 
the electromagnetic coupling the current is, in the rest frame of the 
decaying particle,
\begin{equation}
j^\mu=-\text{Tr}\int\frac{d^4q}{(2\pi)^4}\gamma^\mu\chi(q)= 
-\text{Tr}\int\frac{d{\bf q}}{(2\pi)^{3}} \gamma^\mu \Psi({\bf q}).
\end{equation}
The corresponding Feynman diagram is given in Fig.~\ref{leptonfig}.

Taking the trace we find that $j^0=0$ (whence the current is 
conserved) and
\begin{equation}
{\bf j}=\int\frac{d{\bf q}}{(2\pi)^3} \left(\frac{4m}{q^2}{\bf 
q}\Psi_S(q) Y_{1m}(\hat{\bf q})+4\Psi_V(q) {\bf Y}_{1m}^{(e)}(\hat{\bf 
q})\right). \label{leptonstrom}
\end{equation}
We have here used the expansion (\ref{ettppexp}) of the $1^{--}$~state 
given in Appendix~\ref{waveexp}.

Performing the angular integration we get
\begin{equation}
|{\bf j}|^2=\left( \int\frac{dq}{(2\pi)^{3}} 
\left(8mq\sqrt{\frac{\pi}{3}}\Psi_S(q)+8q^2\sqrt{\frac{2\pi}{3}} 
\Psi_V(q)\right)\right)^2.
\end{equation}

The expression for the leptonic decay is divergent and has to be 
renormalized~\cite{Ito_renorm}.  This can in principle be done in 
several ways.  However, here we must take into account the fact that 
the theory is not fully defined.  This means that with an interaction 
kernel which is instantaneous and defined by the ladder expansion 
only, the Ward-Takahashi identities are not appropriate.  The 
divergence of the wavefunction at the origin can therefore only be 
related to the (experimental) value of the vertex function at some 
suitable momentum transfer.  For non-confining potentials, such as the 
Coulomb potential, there is a region of production of asymptotically 
free fermions above threshold.  The production cross section can then 
be calculated with the kernel and the result be taken as a measure of 
the vertex function.  This is more or less the choice in 
Ref.\cite{Ito_renorm}.  In our case the potential is confining and 
consists of two parts, one of which has a dimensionful coupling 
constant $\lambda$.  This means that the renormalization has to be 
done in the deep inelastic region, where the quarks are asymptotically 
free and the linear part of the potential is negligible.  This 
presents several problems since the lack of manifest covariance makes 
the renormalization constant depend upon both the cut-off and the 
momentum.

The problem appears as follows.  We would like to define a 
renormalization constant $Z_1$, such that
 \begin{equation}
\lim_{\Lambda\to\infty}Z_1(\Lambda)\int^\Lambda d{\bf 
q}\,\gamma^\mu\Psi({\bf q}) =\text{finite.}
\end{equation}
Asymptotically $\Psi({\bf q})\sim q^{-3+\beta}$, and thus 
$j^\mu\sim\Lambda^\beta$ for large $\Lambda$, where $\beta$ is 
determined from the dynamics\cite{Ito_renorm}.

The renormalization constant, $Z_1$, is defined through the 
(analytically continued) vertex function $\Gamma^\mu$ by
\begin{equation}
\bar u({\bf q})\Gamma^\mu u({\bf q})=Z_1^{-1}(q)\bar u({\bf 
q})\gamma^\mu u({\bf q})
\label{zdef}
\end{equation}
and the vertex function satisfies
\begin{eqnarray}
\bbox{\Gamma}({\bf 
q})=&&\bbox{\gamma}+i\int^\Lambda\frac{d^{4}p}{(2\pi)^{4}} K^V({\bf 
p}-{\bf q})\gamma^\mu\frac{1}{\rlap{/}{p}-m} \bbox{\Gamma}({\bf 
p})\frac{1}{\rlap{/}{p}-m}\gamma_\mu\nonumber\\
&&{}+i\int^\Lambda\frac{d^{4}p}{(2\pi)^{4}} K^S({\bf p}-{\bf 
q})\frac{1}{\rlap{/}{p}-m} \bbox{\Gamma}({\bf p})\frac{1}{\rlap{/}{p}-m}.
\end{eqnarray}

The renormalization constant $Z_1$ has the correct asymptotic behavior 
to cancel the $\Lambda$-dependence of the current.  However, due to 
the instantaneous approximation, $Z_1$ depends on $q$.  With only a Coulomb 
potential, $Z_1$ has the correct $\Lambda$-dependence to cancel the 
divergence of $j^{\mu}$ as $\Lambda\to\infty$ for any momentum $q$.  
However, with a confining part added to the potential this would only 
be true in the region $q\to\infty$, where the linear potential is 
negligible.  Unfortunately, $Z_1\sim q^\beta$ when $q\to\infty$, and 
therefore the theory is not renormalizable in this way.

The electromagnetic theory can only be used in lowest order due to the 
non-existence of Ward-Takahashi identities in the instantaneous 
approximation since the theory is not fully defined.  Nevertheless 
there is a loop contribution in the current $j^{\mu}$ which has to be 
renormalized.  As the vertex function cannot be renormalized, we therefore 
adopt the philosophy that, for confining potentials, the most natural 
experimental datum at which to fix the normalization is the resonance 
formation itself.  The renormalization should therefore be effected by 
using, say, the decay width for the lowest resonance as input.  The 
other decay widths are then calculated relative to this.  Thus we will 
optimize only the ratios of the decay constants for successive 
$nS$~states as predictable by the model.  These ratios are independent 
of the cut-off $\Lambda$ for $\Lambda \geq 30 $ GeV to within $1 \%$.
	 
Another way to avoid the problem with infinities would be to redefine 
the potential to take a finite value for $r=0$.  This would make the 
current (\ref{leptonstrom}) finite.  However, the theory must still be 
renormalized although only a finite renormalization is necessary in 
this case.  Therefore such an approach would still face the same 
problem as explained above, as a new dimensionful constant must be 
introduced in the cut-off of the potential.  Our opinion is that our 
approach is much cleaner and conceptually more transparent.

\section{$\gamma\gamma$ decays} 
\label{ggdecay}
The calculation of the $\gamma\gamma$ decays is most easily performed in the 
rest frame of the bound state, which decays into two photons with 
momenta $k_1=(M/2,{\bf k})$ and $k_2=(M/2,-{\bf k})$.  To lowest order 
the matrix element is given by
\begin{eqnarray}
M^{\mu\nu}=&&\text{Tr}\int\frac{d^{4}q}{(2\pi)^{4}} 
\chi(q)\left(\gamma^\mu \frac{1}{\rlap{/}{P}/2+\rlap{/}{q}-\rlap{/}{k_1}-m} 
\gamma^\nu+\gamma^\nu \frac{1}{\rlap{/}{P}/2+\rlap{/}{q}-\rlap{/}{k_2}-m} 
\gamma^\mu\right)\nonumber\\
=&&\text{Tr}\int\frac{d^{4}q}{(2\pi)^{4}} \chi(q)\left(\gamma^\mu 
\frac{1}{\rlap{/}{q}+\bbox{\gamma}\cdot{\bf k}-m} \gamma^\nu+\gamma^\nu 
\frac{1}{\rlap{/}{q}-\bbox{\gamma}\cdot{\bf k}-m} \gamma^\mu\right).
\end{eqnarray}
The corresponding Feynman diagrams are given in Fig.~\ref{gamgamfig}.  
The decay probability is given by
\begin{equation}
\Gamma=\frac{3\pi\alpha^2e_q^4}{2(2J+1)M}\sum|\epsilon_1^\mu\epsilon_2^\nu 
M_{\mu\nu}|^2,
\end{equation}
where $e_q$ is the charge of the quark, and $M$ is the meson mass and 
$J$ its angular momentum.

The amplitude $\epsilon_1^\mu\epsilon_2^\nu M_{\mu\nu}$ can be 
expressed in the scalar amplitudes for the different decaying states.  
This is discussed in Appendix~\ref{ggdecays}.

\section{M1 and E1 transitions} 
\label{m1e1decay}
In the rest frame of the initial particle the matrix element for the 
M1 and E1 transitions is, in the ladder approximation, given by
\begin{eqnarray}
	M^\mu= & & 
	\text{Tr}\int\frac{d^4q}{(2\pi)^4}\bar\chi_{f}(q-k/2,P_f) 
	\gamma^\mu\chi_i(q,M_i) (-\frac{M_i}{2}\gamma^0+\rlap{/}{q}-m)\nonumber 
	\\
	 & & +\text{Tr}\int\frac{d^4q}{(2\pi)^4}\bar\chi_{f}(q+k/2,P_f) 
	 (\frac{M_i}{2}\gamma^0+\rlap{/}{q}-m)\chi_i(q,M_i) \gamma^\mu.
\end{eqnarray}

Here $\chi_i(q,M_i)$ is the wave function of the initial particle in 
its rest frame, and $\bar\chi_f(q\pm k/2,P_f)$ is the wave function of 
the final particle in a frame where it moves with momentum $P_f$.  
This wave function is found by Lorentz transforming the one in the 
rest frame, see Appendix~\ref{lorentz}.  The Feynman diagrams 
corresponding to $M^\mu$ are given in Fig.~\ref{m1fig}.

The decay probability is given by
\begin{equation}
\Gamma=\frac{\alpha e_q^2 k}{2M_i^2(2J+1)}\sum|\epsilon_\mu M^\mu|^2,
\end{equation}
where $e_q$ is the charge of the quark, $k$ the photon momentum, $M_i$ 
the mass of the initial meson and $J$ its angular momentum.

To calculate the matrix element we use the expansions of the wave 
functions given in Appendix~\ref{amplitudes}.  We first perform the 
$q^0$ integration.  Since the Lorentz transformed state 
$\chi(q-k/2,P_f)$ has a very complicated analytic structure we cannot 
make a simple contour integration.  We have to use the principal value 
technique.  The real part is given by a sum of residues
\begin{equation}
\text{Re}\ i\int dq^0 \frac{f(q^0)}{g(q^0)\pm i\epsilon}=\pm \pi 
\frac{f(p^0_k)}{|g'(p^0_k)|},
\end{equation}
where $p^0_k$ is a zero of $g(p^0)$.  The imaginary part gives an 
unphysical non-Hermitian contribution to the decay width.  It should 
vanish in any consistent covariant theory, but in the Salpeter model 
is has been verified that it does not vanish\cite{ResagochMunz4}, 
because of the lack of manifest covariance.  However, for the 
charmonium system this unphysical contribution is negligible, and 
we therefore neglect it.

The integration over the azimuthal angle $\varphi$ can also be done 
analytically.  The rest of the calculation is made numerically.  
Trying to take the trace analytically creates enormously complicated 
expressions which are hardly of any use.

\subsection{Gauge invariance of the matrix element}
In a Bethe-Salpeter model without the instantaneous approximation 
where we have an even kernel of the form $K(p-q)$, the gauge 
invariance of the matrix element can be shown with the help of the 
Bethe-Salpeter equation.  Let the initial meson have momentum $P_i$ 
and the final meson have momentum $P_f=P_i-k$, where $k$ is the photon 
momentum.  Then
\begin{eqnarray}
k_\mu M^\mu=&&\int \frac{d^{4}q}{(2\pi)^{4}}\bar\chi(q-k/2,P_f 
)\rlap{/}{k}\chi(q,P_i)(-\frac{\rlap{/}{P}_i}{2}+\rlap{/}{q} -m)\nonumber\\
=&&\int \frac{d^{4}q}{(2\pi)^{4}} 
\bar\chi(q-k/2,P_f)(\frac{\rlap{/}{P}_i}{2}+\rlap{/}{q}-m)\chi(q,P_i) 
(-\frac{\rlap{/}{P}_i}{2}+\rlap{/}{q}-m)\nonumber\\
&&{}-\int \frac{d^{4}q}{(2\pi)^{4}} \bar\chi(q-k/2,P_f 
)(\frac{\rlap{/}{P}_f}{2}+\rlap{/}{q}-\frac{\rlap{/}{k}}{2}-m) \chi(q,P_i 
)(-\frac{\rlap{/}{P}_f}{2}+\rlap{/}{q}-\frac{\rlap{/}{k}}{2}-m)\nonumber\\
=&&\int \frac{d^{4}q}{(2\pi)^{4}}\frac{d^{4}p}{(2\pi)^{4}} 
\bar\chi(q-k/2,P_f)K(q-p)\chi(p,P_i)\nonumber\\
&&{}-\int \frac{d^{4}q}{(2\pi)^{4}}\frac{d^{4}p}{(2\pi)^{4}} 
\bar\chi(p,P_f)K(q-k/2-p)\chi(q,P_i)\nonumber\\
=&&\int \frac{d^{4}q}{(2\pi)^{4}}\frac{d^{4}p}{(2\pi)^{4}} 
\bar\chi(q-k/2,P_f)K(q-p)\chi(p,P_i)\nonumber\\
&&{}-\int \frac{d^{4}q}{(2\pi)^{4}}\frac{d^{4}p}{(2\pi)^{4}} 
\bar\chi(p-k/2,P_f)K(q-p)\chi(q,P_i)=0
\end{eqnarray}
provided the order of integrations can be interchanged.  This proves 
gauge invariance.

However, if the kernel depends on the momentum of the bound state, as 
it does in the instantaneous approximation, the terms in the last line 
do not cancel.  In the first term $K$ depends on $P_i$, and in the 
second term on $P_f$.

Nevertheless, in the instantaneous approximation we have verified that 
${\bf k}\cdot{\bf M}=0$ and $M^0=0$ in the rest frame of the decaying 
particle, by performing the angular integration of the decay 
amplitude.  Thus gauge invariance is fulfilled.

\section{Numerical results} 
\label{resultat}
The eigenvalue equations given in Appendix~\ref{eigenequ} were solved 
numerically by expanding the wave function in cubic Hermite 
splines\cite{hermite}. However, first the integration region $0\le p 
<\infty$ was transformed into the finite interval $-1\le x\le 1$ by 
$x=(p-p_{0})/(p+p_{0})$, where $p_{0}$ was chosen in the physically 
relevant region.  The integration and the following matrix eigenvalue 
problem for the expansion coefficients were solved with standard 
methods.  We have tested the stability of the solutions by varying the 
number of basis functions and the parameter $p_{0}$, and found that 
the eigenvalues do not change.  Problems, like the ones reported in 
Refs.\cite{olsson_stabil,parramore1,parramore2} for lighter quarks,
have not been seen.

The parameters $\bar{\alpha}_{s}$, $m$, $\lambda$, and $U$ of the 
model are determined by making a least square fit to the mass spectrum 
and the ratios of the lepton pair decay widths $J/\Psi/\Psi(2S)$ and 
$\Psi(2S)/\Psi(3S)$.  As discussed in Sec.~\ref{leptondecay} we 
renormalize the decay widths to the experimental value of the decay of 
$J/\Psi$.  We therefore make the fit to the ratios between the widths 
rather than the widths themselves.  We minimize the function
\begin{equation}
\chi^2=\sum\frac{(M_i(\text{theory})-M_i(\text{exp}))^2}{\sigma_i^2}+ 
\sum\frac{(\Gamma_i(\text{theory})/\Gamma_j(\text{theory})
-\Gamma_i(\text{exp})/\Gamma_j(\text{exp}))^2}{\sigma_{i/j}^{2}}.\label{chikvadrat}
\end{equation}
The first sum goes over all the charmonium masses in 
Table~\ref{masstable} except $\eta_c(2S)$ 
and $h_{c1}$ which are not used in the fit because they still need 
experimental confirmation.  The second sum is over the two ratios of 
lepton pair decay widths $\Gamma(J/\Psi)/\Gamma(\Psi(2S))$ and 
$\Gamma(\Psi(2S))/\Gamma(\Psi(3S))$.  We have not used the decay 
widths for the higher states because their wave functions are not good 
enough to be used for calculations of decay widths. The experimental 
errors for the ratios, $\sigma_{i/j}$, are numerically 
$\sigma_{J/\Psi/2S}=0.30$ and $\sigma_{2S/3S}=0.64$ respectively.

As many of the masses are very accurately measured, using the true 
experimental error for the masses in this function would make the 
masses totally dominate over the decay width.  We therefore use a 
minimum value of $\sigma_{i}$ when the experimental error is small.  
Two fits for both gauges have been made, one where we have used 
$\sigma_i=10$~MeV for the masses when the experimental error is 
smaller than 10~MeV, and one where we have used $\sigma_i=50$~MeV for 
the masses when the experimental error is smaller than 50~MeV. We use 
the true experimental errors for $\sigma_{i/j}$ in both fits.  In the 
second fit we thus give the lepton decays a higher weight, though in 
both cases the masses get a higher weight than the decay widths as the 
errors of the decay widths are rather large.  The numerical results of 
the four fits can be found in Tables \ref{masstable} and 
\ref{decaytable}.  In these tables F and C stand for Feynman and 
Coulomb gauge respectively, and 10 and 50 MeV stand for the minimum 
value of $\sigma_{i}$.

Making four fits gives us a chance to see how sensitive the different 
decay observables are to changes of the model parameters.  The 
parameter values in the four different cases can be found in Table 
\ref{parametertable}.  We see that when we change $\sigma_{i}$ from 
10~MeV to 50~MeV the parameter which changes most is the constant $U$.  
One would perhaps expect that $\bar \alpha_{s}$ would be the parameter 
most sensitive to such a change in $\sigma_{i}$, as the lepton pair 
decays should be very sensitive to $\bar \alpha_{s}$ because they 
depend on the value of the wavefunction at the origin.  However, we 
calculate the ratios between the decay widths rather than the widths 
themselves, and these ratios are not very sensitive to the value of 
$\bar \alpha_{s}$.  This can be understood by noting that in a 
non-relativistic model with a Coulomb potential the ratios of the wave 
functions' value at the origin are independent of the coupling 
constant.  On the other hand, $\bar \alpha_{s}$ of course changes a 
lot between the fits in the two different gauges to compensate for the 
differences in the structure of the vector part of the potential, 
while the other parameters remain almost constant.

The bottom row in Table~\ref{masstable} shows the total $\chi^{2}$ for 
the four fits.  A fictitious error of 10~MeV for all masses has been 
used to calculate the values in order to make them comparable.  We see 
that we obtain a slightly better result in the Coulomb gauge and that 
in the Feynman gauge we obtain the largest difference between the fits 
with different $\sigma_{i}$.

In Table~\ref{decaytable} the $\chi^{2}$ values in the row just above 
the lepton pair decays measure the quality of the calculation of the 
E1, M1, and two gamma transitions. We have here excluded the decay 
$\Psi(2S)\to\eta_{c}\gamma$ as it would totally dominate $\chi^{2}$. 
We see that the results in the Coulomb gauge changes most and that 
the Feynman gauge on average gives a better result although the Coulomb 
gauge with $\sigma_{i}=50$~MeV is better than any of the two fits in 
Feynman gauge. 

In the last row in the same table we find the $\chi^{2}$ values for 
the lepton pair decays.  The decays of the D states have been excluded 
in the calculation as they depend on higher order corrections not 
included in our calculations. We see that the Feynman gauge clearly 
gives a better result.

In conclusion the Coulomb gauge gives better results for the masses, 
the Feynman gauge gives better results for the lepton pair decays and the 
results for the two gamma, E1, and M1 transitions are fairly equal, 
perhaps slightly better in Feynman gauge.  The overall differences are 
not so large that we can clearly tell which one of the two gauges 
gives best results.

The lepton decay widths are normalized such that the value for 
$J/\Psi$ agrees with the experimental value.  We see that the other 
values for the S states become too large.  This indicates that we lack 
a fine-tuning of the shape of the potential.  A smaller slope of the 
linear confinement potential would push the wave function further away 
from the origin but would on the other hand make the mass spectrum 
worse.  That the decay widths for the D states are rather small is not 
surprising, as they mainly come from higher order corrections not 
included here.  In non-relativistic models with a Coulomb potential 
the wave function for D states is zero at the origin, which give a 
vanishing decay width.  In the Salpeter model this means that higher 
order corrections to the kernel in the ladder approximation are very 
important for the D states.

The values for the two gamma decays agree very well with experiment, 
and the values for the E1 transitions agree fairly well although the 
values for the decays including the $\chi_{c1}$ state all come out 
about a factor of two too large. 

When it comes to the M1~transitions we see that the predicted value of 
the $J/\Psi\to \gamma\eta_c$ decay width agrees very well with 
experimental data.  On the other hand the $\Psi(2S)$ decay width is 
far too large.  This is because the wave function of the excited state 
is not determined well enough, and as this decay is forbidden it is 
extremely sensitive to the exact shape of the wave functions.

For light quarks, the non-relativistic quark model parameters 
usually relate the magnetic moments directly to the constituent masses 
and there is no room in the parameterization, so to speak, to 
introduce anomalous magnetic moments.

For the charm quark the current and constituent masses should be 
essentially the same and there is then a possibility to introduce an 
anomalous magnetic moment as a new independent parameter.

The details of the decay mechanism are in principle sensitive to this 
parameter.  In the absence so far of any measurement of the magnetic 
moment of charmed baryons, the radiative decay rates therefore 
constitute at present the only place where this parameter can be 
studied.

Once an accurate determination of the wave functions can be made, the 
anomalous magnetic moment can be calculated from the M1 transitions. 
But it is important to notice that the outcome of such a calculation 
depends heavily on the quark mass. Therefore a determination of 
the quark mass must be made first. An estimate of the anomalous 
magnetic moment have previously been made in a Salpeter model where a 
non-relativistic approximation has been made\cite{Hulth_o_Hakan}.

A calculation of lepton pair decays and E1 transitions has earlier 
been done by Ito\cite{Ito_E1} in a model where only the positive energy 
term is used in the Salpeter equation. Recently a calculation of all 
decay observables has been done by Resag et al.\cite{ResagochMunz3} in 
a Salpeter model, but with a cut-off in the potential to regularize 
the divergence in the lepton pair decay amplitude. The overall results 
in both these calculations are of about the same quality as in ours. 
As we have discussed earlier, the lepton pair decays depend 
of the cut-off parameter, and therefore have to be renormalized before 
they can be compared. Doing so, we see that their results for a 
scalar confinement are almost identical with our results in Feynman 
gauge with $\sigma_{i}=10$~MeV.

\section{Importance of non-dominant terms}
Sometimes the Salpeter equation is reduced to only the term with the 
positive energy states of the quark. This is done by making the 
following replacement of the propagators (\ref{propagator}) in the 
Salpeter equation:
\begin{equation}
	\frac{1}{\rlap{/}{q}-m} \longrightarrow 
	\frac{\Lambda_{+}^{+}({\bf q})\gamma^{0}}{(q^{0}-E({\bf 
	q})+i\epsilon)}= \frac{\gamma^{0}\Lambda_{-}^{-}({\bf q})}{(q^0+E({\bf 
	q})-i\epsilon)}
\end{equation}
This imposes an extra constraint on the reduced wave function $\Psi({\bf 
q})$. Eq.~(\ref{litet_psi}) is replaced by
\begin{equation}
	\Lambda_{-}^{+}({\bf q})\Psi({\bf q})\Lambda_{+}^{-}({\bf q})=0
\end{equation}
For the wave function $\chi(q)$ only the first term in (\ref{waveeq}) 
remains:
\begin{equation}
	\chi(q)=-\frac{i(M-2E)}{(M/2+q^0-E+i\epsilon)(-M/2+q^0+E-i\epsilon)} 
	\Lambda_{+}^{+}({\bf q})\Psi({\bf q})\Lambda_{-}^{-}({\bf q})
\end{equation}

Does this reduction change the results significantly?  Olsson {\it et 
al.\/}\cite{olsson_2} have recently investigated the validity of the 
reduced Salpeter equation by comparing the eigenvalues and wave 
functions found from the full and reduced Salpeter equation.  They 
found that there is only a significant difference for light quarks, 
and for as heavy ones as the charm quark the difference is very small.

However, this does not completely test the validity of the reduced 
Salpeter equation.  In particular this does not tell us how important 
the last two terms in (\ref{waveeq}) are.  They do not appear in the 
calculation of the mass spectrum, which only involves the reduced wave 
function $\Psi({\bf q})$.  On the other hand, the E1 and M1 
transitions do include these two terms.

To test the validity of the reduced Salpeter equation we have 
therefore compared the decay widths for E1 and M1 transitions using only the 
terms with positive energy states and the widths using all terms. We 
have not solved the reduced Salpeter equation but only neglected all 
but the dominant term. This will not give us an exact comparison with 
the reduced Salpeter equation but as have been shown by Olsson~{\it et 
al.\/}\cite{olsson_2} the wave functions differ very little. The
result is shown in Table \ref{positiveenergytable}.  We see that the 
decay widths change considerably if we neglect all but the dominant 
term, in some cases with more than 100~\%.  This
shows the importance of keeping all terms.  Although the magnitude of 
the negative energy part of the reduced wave functions is small -- the 
contribution to the norm is typically $<5$~\% -- there are
other terms in the decay amplitudes which make the contribution from 
the non-dominant terms significant.

This can be understood if we note that the last two terms in the wave 
function (\ref{waveeq}) do not appear in in the expression for the 
norm (\ref{normtr}).  Using the reduced Salpeter equation leads to a 
normalization condition with only the first term in (\ref{normtr}).  
The omitted second term is small and thus changes the normalization 
very little.

To further illustrate this we have plotted the functions
\begin{displaymath}
	\frac{m}{E(q)}\Psi_{P}(q)\pm \Psi_{A}(q)
\end{displaymath}
and
\begin{displaymath}
	\frac{I_{0}(q)}{E(q)}
\end{displaymath}
for the $0^{-+}$ state in Fig.~\ref{amplitudplot}. These functions 
appears in the amplitudes of $0^{+-}$'s  wave function. We see that 
negative energy amplitude $\frac{m}{E(q)}\Psi_{P}(q)- \Psi_{A}(q)$ is 
considerably smaller than the positive energy amplitude. The 
amplitude $I_{0}/E$ appearing in the last two terms of the wave 
function (\ref{waveeq}) is somewhere in between in magnitude. However, 
it should be noted that it is difficult to directly compare this 
function to the other two because of the other factors in the 
amplitudes coming from the propagators. But we see that it is clearly 
not negligible.

Comparing the mass spectrum and wave functions is therefore not enough 
to tell if an approximation is valid.  The decay widths must also be 
studied.  Using the reduced Salpeter equation means that several terms 
in the decay widths are omitted and we have shown that these terms are 
very important in the E1 and M1 transitions.  We conclude that the 
reduced Salpeter equation gives approximately the same result as the 
full Salpeter equations only for those quantities which do not depend 
on the last two terms in the wave function (\ref{waveeq}).

\section{Summary and conclusions} 
\label{tillsist}
We have done a thorough analysis of the charmonium system in a 
Salpeter model with an instantaneous potential.  We have presented 
general equations for calculating the mass spectrum as well as decay 
widths for lepton pair decays, two gamma decays, M1 and E1 
transitions.  This has been done for a vector plus scalar potential 
both in the Coulomb and Feynman gauges.  The formalism used is based 
on the one developed by Suttorp\cite{Suttorp}.

We have discussed the problem with the renormalization of the lepton 
pair decay amplitude in the instantaneous approximation.  This 
approximation makes the renormalization constant $Z_{1}$ depend on 
$q$ and $Z_{1}\to\infty$ as $q\to\infty$.  With a linear confining 
potential we have to make the renormalization in the region where the 
quarks are asymptotically free and the linear potential is negligible.  
But since $Z_{1}\to\infty$ this leads to an inconsistency, and 
therefore the only natural way to define the renormalization constant 
is to normalize the calculated widths to the experimental value of the 
width $J/\Psi\to e^{+}e^{-}$.

Numerically we have obtained an excellent fit to the mass spectrum of 
charmonium.  Most of the decay widths agree rather well with 
experiments, although the lepton pair decays of the excited S states 
are too large.  The excited states are obviously more sensitive to 
fine-tuning of the model.  A smaller slope of the linear confinement 
potential would push the wave function further away from the origin 
but would on the other hand make the mass spectrum worse.  Similarly 
the poor result for the forbidden M1 transition 
$\Psi(2S)\to\eta_{c}\gamma$ indicates a lack of fine-tuning of the 
potential.  We conclude that before any detailed predictions can be 
made from the Salpeter equation, a systematic way to improve the 
kernel must be developed.

Comparing the two gauges, Feynman and Coulomb gauge, we have found the 
the Coulomb gauge gives better results for the mass spectrum and that 
the Feynman gauge gives better results for the lepton pair decays. 
The overall difference is not large enough to clearly tell which 
gauge gives the best results.

The validity of the reduced Salpeter equation has been tested, and it 
has been found that the decay widths of the E1 and M1 transitions 
change considerably if one uses the reduced equation.

\acknowledgments
This work was supported by the Swedish Natural Science Research 
Council (NFR), contract F-AA/FU03281-308.

\appendix

\section{Wave functions for different states} 
\label{waveexp}

We here present the expansion in the Dirac algebra of the reduced wave 
functions, $\Psi({\bf q})$, defined in (\ref{redwave}), for different 
states.  These expansions are the most general ones with the correct 
transformation properties regarding angular momentum, parity, and 
charge parity, which also fulfill the conditions 
(\ref{psieq3}-\ref{psieq4}).  All scalar functions $\Psi_i$ depend 
only on $q=|{\bf q}|$.

\begin{eqnarray}
0^{-+}:&&\quad \Psi({\bf 
q})=\left(\Psi_P(q)\gamma^5+\Psi_A(q)\left(\gamma^0\gamma^5+ 
\gamma^5\frac{\bbox{\alpha}\cdot{\bf 
q}}{m}\right)\right)Y_{00}(\hat{\bf q}),\\
1^{--}:&&\quad \Psi({\bf 
q})=\left(\Psi_S(q)\left(1+\frac{m}{q^2}\bbox{\gamma}\cdot{\bf 
q}\right)+ \Psi_{T_1}(q)\frac{\bbox{\alpha}\cdot{\bf 
q}}{q}\right)Y_{1m}(\hat{\bf q})\nonumber\\
&&\quad{}+\left(\Psi_{V}(q)\bbox{\gamma}+\Psi_{T_2}(q)
\left(\bbox{\alpha}-i\gamma^5 
\frac{{\bf q}\times\bbox{\gamma}}{m}\right)\right)\cdot{\bf 
Y}_{1m}^{(e)}(\hat{\bf q}), \label{ettppexp}\\
1^{+-}:&&\quad \Psi({\bf 
q})=\left(\Psi_P(q)\gamma^5+\Psi_A(q)\left(\gamma^0\gamma^5+ 
\gamma^5\frac{\bbox{\alpha}\cdot{\bf 
q}}{m}\right)\right)Y_{1m}(\hat{\bf q}),\\
0^{++}:&&\quad \Psi({\bf 
q})=\left(\Psi_S(q)\left(1+\frac{m}{q^2}\bbox{\gamma}\cdot{\bf 
q}\right)+ \Psi_{T}(q)\frac{\bbox{\alpha}\cdot{\bf 
q}}{q}\right)Y_{00}(\hat{\bf q}),\\
1^{++}:&&\quad \Psi(q)=\left(\Psi_V(q)i\frac{({\bf 
q}\times\bbox{\gamma})}{q}+\Psi_T(q)\left(i 
\frac{(\bbox{\alpha}\times{\bf 
q})}{q}+\frac{q}{m}\gamma^5\bbox{\gamma}\right)\right) \cdot{\bf 
Y}^{(e)}_{1m}(\hat{\bf q}),\\
2^{++}:&&\quad \Psi({\bf 
q})=\left(\Psi_S(q)\left(1+\frac{m}{q^2}\bbox{\gamma}\cdot{\bf 
q}\right)+ \Psi_{T_1}(q)\frac{\bbox{\alpha}\cdot{\bf 
q}}{q}\right)Y_{2m}(\hat{\bf q})\nonumber\\
&&\quad{}+\left(\Psi_{V}(q)\bbox{\gamma}+\Psi_{T_2}(q)
\left(\bbox{\alpha}-i\gamma^5 
\frac{{\bf q}\times\bbox{\gamma}}{m}\right)\right)\cdot{\bf 
Y}_{2m}^{(e)}(\hat{\bf q}).
\end{eqnarray}
Here the $Y_{\ell m}(\hat{\bf q})$'s are the scalar spherical 
harmonics and the ${\bf Y}_{\ell m}^{(e)}(\hat{\bf q})$'s are the 
electric vector spherical harmonics.  The latter are defined in terms 
of the vector spherical harmonics,
\begin{equation}
	{\bf Y}_{JL}^{m}=\sum_{m_{1}m_{2}} Y_{Lm_{1}}{\bf e}_{m_{2}}(Lm_{1}\ 
	1m_{2}|Jm),
\end{equation}
where ${\bf e}_{\pm}=\mp({\bf e}_{x}\pm i{\bf e}_{y})/\sqrt{2}$ and ${\bf 
e}_{0}= {\bf e}_{z}$, as
\begin{equation}
	{\bf Y}_{Jm}^{(e)}=\sqrt{\frac{J+1}{2J+1}}{\bf Y}_{J,J-1}^{m} + 
	\sqrt{\frac{J}{2J+1}}{\bf Y}_{J,J+1}^{m}.
\end{equation}

\section{Normalization condition}
The wave functions satisfy the normalization condition~\cite{Lurie}
\begin{equation}
\text{Tr}\int\frac{d^4p}{(2\pi)^4}\int\frac{d^4q}{(2\pi)^4} 
\bar\chi(p)P^\mu\frac{d}{dP^\mu}(I(p,q,P)+iK(p,q,P))\chi(q)=2iM^2,
\label{normcond}
\end{equation}
where
\begin{equation}
I(p,q,P)=(2\pi)^4\delta^{(4)}(p-q)S_1^{-1}(P/2+p)S_2^{-1}(-P/2+p).
\end{equation}
The adjoint wave function $\bar\chi$ is defined
\begin{equation}
\bar\chi=\gamma^0\chi^+\gamma^0.
\end{equation}
We have contracted with $P^\mu$ in (\ref{normcond}).  This is possible 
if the Salpeter equation is written in a covariant form.  This is done 
by letting the kernel depend on the four-vectors $(0,{\bf q})$ and 
$(0,{\bf p})$ in the rest frame.  The Salpeter equation is then Lorentz 
transformed by first multiplying it by the spinor transformation 
matrices $S(\Lambda)$ and $S^{-1}(\Lambda)$ from each side.  Then the 
change of variables $q\to \Lambda^{-1}q$ is made.  The kernel is found 
to depend on the orthogonal part of the momentum 
$q_\perp=q-\frac{q\cdot P}{M^2}P$ and $p_\perp=p-\frac{p\cdot 
P}{M^2}P$ in a frame where the particle moves with momentum $P$.  
In the rest frame $q_\perp$ reduces to $(0,{\bf q})$.  Thus 
$K=K(q_\perp-p_\perp)$.  For such kernels we have
\begin{equation}
P^\mu\frac{d}{dP^\mu}K(q_\perp-p_\perp)=0.
\end{equation}
Thus the second term in (\ref{normcond}) does not contribute to the 
normalization.  Integrating the other term over $p$ gives the 
following normalization condition in the rest frame,
\begin{equation}
\mbox{Tr}\int\frac{d^4q}{(2\pi)^4}\bar\chi(q)\gamma^0\chi(q) 
(-\frac{M}{2}\gamma^0+\rlap{/}{q}-m)=2iM.
\end{equation}
Using the form (\ref{waveeq}) of the wave function and integrating 
over $q^0$ gives
\begin{equation}
\text{Tr}\int\frac{d{\bf q}}{(2\pi)^{3}}\left(\Psi^+({\bf 
q})\Lambda_{+}^{+}({\bf q}) \Psi({\bf q})\Lambda_{-}^{-}({\bf q}) 
-\Psi^+({\bf q})\Lambda_{-}^{+}({\bf q})\Psi({\bf 
q})\Lambda_{+}^{-}({\bf q})\right)=2M.\label{normtr}
\end{equation}
Note that due to the covariant formulation of the normalization 
condition, the wave function is correctly normalized in any frame.

By taking the trace in (\ref{normtr}) for the different states we 
find the following normalization conditions:
\begin{eqnarray}
0^{-+}:&&\quad 
2M=\frac{8}{m}\int_0^\infty\frac{dq}{(2\pi)^3}q^2E(q)\Psi_A(q)\Psi_P(q),\\
1^{--}:&&\quad 2M=8\int_0^\infty \frac{dq}{(2\pi)^{3}} 
q^2\left(\frac{E(q)}{q}\Psi_S(q)\Psi_{T_1}(q) 
+\frac{E(q)}{m}\Psi_{T_2}(q)\Psi_V(q)\right),\\
1^{+-}:&&\quad 
2M=\frac{8}{m}\int_0^\infty\frac{dq}{(2\pi)^3}q^2E(q)\Psi_A(q)\Psi_P(q),\\
0^{++}:&&\quad 2M=8\int_0^\infty \frac{dq}{(2\pi)^3} 
qE(q)\Psi_S(q)\Psi_{T}(q),\\
1^{++}:&&\quad 2M=-8\int_0^\infty \frac{dq}{(2\pi)^{3}} 
q^2\frac{E(q)}{m}\Psi_V(q)\Psi_T(q),\\
2^{++}:&&\quad 2M=8\int_0^\infty \frac{dq}{(2\pi)^{3}} 
q^2\left(\frac{E(q)}{q}\Psi_S(q)\Psi_{T_1}(q) 
+\frac{E(q)}{m}\Psi_{T_2}(q)\Psi_V(q)\right).
\end{eqnarray}

\section{Eigenvalue equations for the different states} 
\label{eigenequ}
Equations for the scalar amplitudes of the different states can be 
found from (\ref{psiequation}).  By expanding both sides of 
(\ref{psiequation}) in the Dirac algebra, and then multiplying by 
$(Y_{\ell m}\hat{\bf q})^*$ or $({\bf Y}_{\ell m}^{(e)}(\hat {\bf 
q}))^*$ and summing over $m$, and finally integrating over the angular 
variables we get the equations below\cite{Suttorp}.  The 
functions $F_\ell^V(p,q)$ and $F^S_\ell(p,q)$ are the coefficients of 
the spherical harmonics in the expansion of the kernel given in 
Section~\ref{potential}. The function $F_{\ell}^{Vc}(p,q)$ is the 
same as $F_\ell^V(p,q)$, but the terms with $F_{\ell}^{Vc}(p,q)$ only 
appear in the Coulomb gauge. Thus, in the Feynman gauge, these 
terms should be set to zero.

\subsection*{$0^{-+}$~state}
\begin{eqnarray}
M\Psi_P(q)=&&\left(\frac{2E(q)^2}{m} + U\frac{m}{E(q)}\left(1- 
\frac{q^{2}}{m^{2}}\right)\right)\Psi_A(q)\nonumber\\
&&{}+\frac{m}{E(q)}\int_0^\infty 
\frac{dp}{(2\pi)^{3}} 
p^2\left(F_0^S(p,q)-\frac{pq}{m^2}F_1^S(p,q)\right)\Psi_A(p)\nonumber\\
&&{}+\frac{m}{E(q)}\int_0^\infty\frac{dp}{(2\pi)^{3}} p^2 
\left(2F_0^V(p,q)-F_0^{Vc}(p,q)-\frac{pq}{m^{2}}F_{1}^{Vc}(p,q)\right)\Psi_A(p),\\
M\Psi_A(q)=&&m\left(2 + \frac{U}{E(q)}\right)\Psi_P(q)\nonumber\\
&&{}+\frac{m}{E(q)}\int_0^\infty\frac{dp}{(2\pi)^{3}} 
p^2 \bigl(F_0^S(p,q)-4F_0^V(p,q)+F_{0}^{Vc}(p,q)\bigr)\Psi_P(p).
\end{eqnarray}

\subsection*{$1^{--}$~state}
\begin{eqnarray}
\lefteqn{M\Psi_{T_1}(q)=\left(\frac{2E(q)^2}{q} + 
\frac{U}{qE(q)}(m^{2}-q^{2})\right)\Psi_S(q)}\nonumber\\
&&{}-\frac{1}{E(q)}\int_0^\infty\frac{dp}{(2\pi)^{3}} 
p^2\left(\left(qF_1^S-\frac{m^2}{3p} 
(F_0^S+2F_2^S)\right)\Psi_S(p)-m\frac{\sqrt{2}}{3}(F_0^S-F_2^S)\Psi_V(p) 
\right)\nonumber\\
&&{}-\frac{1}{E(q)}\int_0^\infty\frac{dp}{(2\pi)^{3}} p^2 
\Biggl(\left(q(4F_1^V-F_{1}^{Vc})+\frac{m^2}{3p} 
(2F_0^V+F_{0}^{Vc}+4(F_2^V-F_{2}^{Vc}))\right) 
\Psi_S(p) \nonumber\\
&&\qquad +\left(m\frac{2\sqrt{2}}{3}(F_0^V-F_2^V) - 
\frac{m(p^{2}-q^{2})\sqrt{2}}{2pq}F_{1}^{Vc}\right)\Psi_V(p) \Biggr),\\
\lefteqn{M\Psi_{T_2}(q)=m\left(2+ \frac{U}{E(q)}\right)\Psi_V(q)}\nonumber\\
&&{}+\frac{1}{E(q)}\int_0^\infty\frac{dp}{(2\pi)^{3}} 
p^2\left(\frac{\sqrt{2}m^2}{3p} 
(F_0^S-F_2^S)\Psi_S(p)+\frac{m}{3}(2F_0^S+F_2^S)\Psi_V(p)\right)\nonumber\\
&&{}-\frac{1}{E(q)}\int_0^\infty\frac{dp}{(2\pi)^{3}} 
p^2\Biggl(\frac{\sqrt{2}m^2}{p} 
\left(\frac{2}{3}(F_0^V-F_2^V)+\frac{(p^{2}-q^{2})}{2pq}F_{1}^{Vc} 
\right)\Psi_S(p) \nonumber\\
&&\qquad+\frac{2m}{3}(2F_0^V-F_{0}^{Vc}+F_2^V+F_{2}^{Vc})
\Psi_V(p)\Biggr),\\
\lefteqn{M\Psi_S(q)=q\left(2+ \frac{U}{E(q)}\right)\Psi_{T_1}(q)}\nonumber\\
&&{}+\frac{1}{E(q)}\int_0^\infty\frac{dp}{(2\pi)^{3}} 
p^2\left(\frac{q}{3}(F_0^S+2F_2^S) 
\Psi_{T_1}(p)+\frac{\sqrt{2}q}{3}(F_0^S-F_2^S)\Psi_{T_2}(p)\right) \nonumber\\
&&{}+\frac{1}{E(q)}\int_{0}^{\infty}\frac{dp}{(2\pi)^{3}} 
p^{2}\left(\frac{p}{3}(F_{0}^{Vc}-4F_{2}^{Vc})\Psi_{T_{1}}(p) 
-\frac{(p^{2}-q^{2})\sqrt{2}}{2p}F_{1}^{Vc}\Psi_{T_{2}}(p)\right), \\
\lefteqn{M\Psi_V(q)=\left(\frac{2E(q)^2}{m} + U\frac{m}{E(q)}\left(1- 
\frac{q^{2}}{m^{2}}\right)\right)\Psi_{T_2}(q)}\nonumber\\
&&{}+\frac{1}{E(q)}\int_0^\infty\frac{dp}{(2\pi)^{3}} 
p^2\left(\frac{\sqrt{2}m}{3} 
(F_0^S-F_2^S)\Psi_{T_1}(p)+\left(\frac{m}{3}(2F_0^S+F_2^S)-\frac{pq}{m} 
F_1^S\right)\Psi_{T_2}(p)\right)\nonumber\\
&&{}+\frac{1}{E(q)}\int_0^\infty \frac{dp}{(2\pi)^{3}} 
p^2\left(\frac{m(p^{2}-q^{2})\sqrt{2}}{2q^{2}}F_{1}^{Vc}\Psi_{T_{1}}(p)
-\left(\frac{2pq}{m}F_1^V+\frac{2m}{3}(F_{0}^{Vc}-F_{2}^{Vc})\right)
\Psi_{T_2}(p)\right).
\end{eqnarray}

\subsection*{$1^{+-}$~state}
\begin{eqnarray}
M\Psi_P(q)=&&\left(\frac{2E(q)^2}{m}+ U 
\frac{m}{E(q)}\left(1-\frac{q^{2}}{m^{2}}\right)
\right)\Psi_A(q)\nonumber\\
&&{}+\frac{m}{E(q)}\int_0^\infty 
\frac{dp}{(2\pi)^{3}} 
p^2\left(F_1^S-\frac{pq}{3m^2}(F_0^S+2F_2^S)\right)\Psi_A(p)\nonumber\\
&&{}+\frac{m}{E(q)}\int_0^\infty\frac{dp}{(2\pi)^{3}} p^2 
\left(2F_1^V-F_{1}^{Vc}+\frac{pq}{3m^{2}}(F_{0}^{Vc}-4F_{2}^{Vc})
\right)\Psi_A(p),\\
M\Psi_A(q)=&&m\left(2+ \frac{U}{E(q)}\right)\Psi_P(q)\nonumber\\
&&{}+\frac{m}{E(q)}\int_0^\infty\frac{dp}{(2\pi)^{3}} 
p^2 \bigl(F_1^S-4F_1^V+F_{1}^{Vc}\bigr)\Psi_P(p).
\end{eqnarray}

\subsection*{$0^{++}$~state}
\begin{eqnarray}
\lefteqn{M\Psi_{T}(q)=\left(\frac{2E(q)^2}{q}+ 
\frac{U}{qE(q)}(m^{2}-q^{2})\right)\Psi_S(q)}\nonumber\\
&&{}-\frac{1}{E(q)}\int_0^\infty\frac{dp}{(2\pi)^{3}} 
p^2\left(qF_0^S-\frac{m^2}{p} F_1^S\right)\Psi_S(p)\nonumber\\
&&{}-\frac{1}{E(q)}\int_0^\infty\frac{dp}{(2\pi)^{3}} p^2 
\left(q(4F_0^V-F_{0}^{Vc})+\frac{m^2}{p} (2F_1^V-F_{1}^{Vc})\right)\Psi_S(p),\\
\lefteqn{M\Psi_S(q)=q\left(2+ \frac{U}{E(q)}\right)\Psi_{T}(q)}\nonumber\\
&&{}+\frac{1}{E(q)}\int_0^\infty\frac{dp}{(2\pi)^{3}} p^2q 
(F_1^S -F_{1}^{Vc})
\Psi_{T}(p).
\end{eqnarray}

\subsection*{$1^{++}$~state}
\begin{eqnarray}
\lefteqn{M\Psi_T(q)=-m\left(2+\frac{U}{E(q)}\right)\Psi_V(q)}\nonumber\\
&&{}-\frac{m}{E(q)}\int_0^\infty\frac{dp}{(2\pi)^{3}} p^2 
(F_1^S-2F_1^V)\Psi_V(p),\\
\lefteqn{M\Psi_V(q)=-\left(\frac{2E(q)^2}{m}+ U\frac{m}{E(q)}\left(1- 
\frac{q^{2}}{m^{2}}\right)\right)\Psi_T(q)}\nonumber\\
&&{}-\frac{m}{E(q)}\int_0^\infty\frac{dp}{(2\pi)^{3}} 
p^2\left(F_1^S-\frac{pq}{3m^2} 
(2F_0^S+F_2^S)\right)\Psi_T(p)\nonumber\\
&&{}+\frac{2}{3E(q)m}\int_0^\infty\frac{dp}{(2\pi)^{3}} 
p^2 \left(pq(2F_0^V+F_2^V)-q^{2}(F_{0}^{Vc}-F_{2}^{Vc})\right)\Psi_T(p).
\end{eqnarray}

\subsection*{$2^{++}$~state}
\begin{eqnarray}
\lefteqn{M\Psi_{T_1}(q)=\left(\frac{2E(q)^2}{q}+ 
\frac{U}{qE(q)}(m^{2}-q^{2})\right)\Psi_S(q)}\nonumber\\
&&{}-\frac{1}{E(q)}\int_0^\infty\frac{dp}{(2\pi)^{3}} 
p^2\left(\left(qF_2^S-\frac{m^2}{5p} 
(2F_1^S+3F_3^S)\right)\Psi_S(p)-m\frac{\sqrt{6}}{5}(F_1^S-F_3^S)\Psi_V(p) 
\right)\nonumber\\
&&{}-\frac{1}{E(q)}\int_0^\infty\frac{dp}{(2\pi)^{3}} p^2 
\Biggl(\left(q(4F_2^V-F_{2}^{Vc})+\frac{m^2}{5p} 
(4(F_1^V+F_{1}^{Vc})+3(2F_3^V-3F_{3}^{Vc}))\right)
\Psi_S(p) \nonumber\\
&&{}+\left(m\frac{2\sqrt{6}}{5}(F_1^V-F_3^V)- 
m\frac{(p^{2}-q^{2})\sqrt{6}}{2pq}F_{2}^{Vc}\right)\Psi_V(p) \Biggr),\\
\lefteqn{M\Psi_{T_2}(q)=m\left(2+ \frac{U}{E(q)}\right)\Psi_V(q)}\nonumber\\
&&{}+\frac{1}{E(q)}\int_0^\infty\frac{dp}{(2\pi)^{3}} 
p^2\left(\frac{\sqrt{6}m^2}{5p} 
(F_1^S-F_3^S)\Psi_S(p)+\frac{m}{5}(3F_1^S+2F_3^S)\Psi_V(p)\right)\nonumber\\
&&{}-\frac{1}{E(q)}\int_0^\infty\frac{dp}{(2\pi)^{3}} 
p^2\Biggl(\left(\frac{2\sqrt{6}m^2}{5p} (F_1^V-F_3^V) 
+\frac{m^{2}(p^{2}-q^{2})\sqrt{6}}{2p^{2}q}F_{2}^{Vc}\right)\Psi_S(p) 
\nonumber \\
&&{}+\frac{2m}{5}(3(F_1^V-F_{1}^{Vc})+(2F_3^V+3F_{3}^{Vc}))\Psi_V(p)\Biggr),\\
\lefteqn{M\Psi_S(q)=q\left(2+\frac{U}{E(q)}\right)\Psi_{T_1}(q)}\nonumber\\
&&{}+\frac{1}{E(q)}\int_0^\infty\frac{dp}{(2\pi)^{3}} 
p^2\left(\frac{q}{5}(2F_1^S+3F_3^S) 
\Psi_{T_1}(p)+\frac{\sqrt{6}q}{5}(F_1^S-F_3^S)\Psi_{T_2}(p)\right)\nonumber\\
&&{}+\frac{1}{E(q)}\int_0^\infty\frac{dp}{(2\pi)^{3}} 
p^{2}\left(\frac{p}{5}(4F_{1}^{Vc}-9F_{3}^{Vc})\Psi_{T_{1}}(p) -
\frac{(p^{2}-q^{2})\sqrt{6}}{2p}F_{2}^{Vc}\Psi_{T_{2}}(p)\right) ,\\
\lefteqn{M\Psi_V(q)=\left(\frac{2E(q)^2}{m}+U\frac{m}{E(q)}\left(1 - 
\frac{q^{2}}{m^{2}}\right)\right)\Psi_{T_2}(q)}\nonumber\\
&&{}+\frac{1}{E(q)}\int_0^\infty\frac{dp}{(2\pi)^{3}} 
p^2\left(\frac{\sqrt{6}m}{5} 
(F_1^S-F_3^S)\Psi_{T_1}(p)+\left(\frac{m}{5}(3F_1^S+2F_3^S)-\frac{pq}{m} 
F_2^S\right)\Psi_{T_2}(p)\right)\nonumber\\
&&{}+\frac{1}{E(q)}\int_0^\infty \frac{dp}{(2\pi)^{3}} 
p^2\left(\frac{(p^{2}-q^{2})\sqrt{6}}{2q^{2}}F_{2}^{Vc} \Psi_{T_{1}}(p) 
-\left(\frac{2pq}{m}F_2^V+\frac{6}{5}(F_{1}^{Vc}-F_{3}^{Vc})
\right)\Psi_{T_2}(p)\right).
\end{eqnarray}

\section{Amplitudes for the different states} 
\label{amplitudes}
For calculation of different decay amplitudes we need the expressions 
for the wave functions $\chi(q)$ for the different states.  Like the 
reduced wave function, $\Psi({\bf q})$, $\chi(q)$ is expanded in the 
Dirac algebra by imposing the correct transformation properties 
regarding angular momentum, parity, and charge parity.  The scalar 
amplitudes of this expansion are then expressed in the amplitudes of 
$\Psi({\bf q})$, by similarly expanding (\ref{waveeq}) in the Dirac 
algebra using the wave functions given in Appendix~\ref{waveexp}.

\subsection*{$0^{-+}$~state}
The most general wave function for the $0^{-+}$~state can be written
\begin{equation}
\chi(q)=(\chi_A(q)\gamma^5+\chi_B(q)\gamma^5(\bbox{\gamma}\cdot{\bf 
q})+ 
\chi_C(q)\gamma^5\gamma^0+\chi_D(q)\gamma^5(\bbox{\alpha}\cdot{\bf q}))
Y_{00}(\hat{\bf q}).
\end{equation}
The scalar amplitudes $\chi_i(q)$ are
\begin{eqnarray}
\chi_A(q)=&&(\Delta^{+-}+\Delta^{-+})\frac{1}{2}\Psi_P(q)+ 
(\Delta^{+-}-\Delta^{-+})\frac{E}{2m}\Psi_A(q), \label{D2}\\
\chi_B(q)=&&(\Delta^{--}-\Delta^{++})\frac{1}{E}I_0(q),\\
\chi_C(q)=&&-(\Delta^{+-}+\Delta^{-+})\frac{1}{2}\Psi_A(q)- 
(\Delta^{+-}-\Delta^{-+})\frac{m}{2E}\Psi_P(q)\nonumber\\
&&+(\Delta^{--}+\Delta^{++})\frac{q^2}{E^2}I_0(q),\\
\chi_D(q)=&&(\Delta^{+-}+\Delta^{-+})\frac{1}{2m}\Psi_A(q)+ 
(\Delta^{+-}-\Delta^{-+})\frac{1}{2E}\Psi_P(q)\nonumber\\
&&+(\Delta^{--}+\Delta^{++})\frac{m}{E^2}I_0(q),
\end{eqnarray}
where
\begin{eqnarray}
\Delta^{+-}=&&-\frac{i(M-2E)}{(M/2+q^0-E+i\epsilon)(-M/2+q^0+E-i\epsilon)} ,
\label{delta1} \\
\Delta^{-+}=&&\frac{i(M+2E)}{(M/2+q^0+E-i\epsilon)(-M/2+q^0-E+i\epsilon)},\\
\Delta^{++}=&&\frac{i}{(M/2+q^0-E+i\epsilon)(-M/2+q^0-E+i\epsilon)},\\
\Delta^{--}=&&\frac{i}{(M/2+q^0+E-i\epsilon)(-M/2+q^0+E-i\epsilon)} .
\label{delta4}
\end{eqnarray}
The integral $I_0(q)$ is
\begin{eqnarray}
I_0(q)=&&U \Psi_{A}(q)+\frac{1}{2}\int\frac{dp}{(2\pi)^3} p^2(F_0^S(p,q)+\frac{p}{q} 
F_1^S(p,q))\Psi_A(p) \nonumber\\
&&\qquad{}+\int\frac{dp}{(2\pi)^3} p^2\left(F_0^V(p,q) 
-\frac{1}{2}\left(F_{0}^{Vc}-\frac{p}{q}F_{1}^{Vc}\right)\right)\Psi_A(p).
\end{eqnarray}

\subsection*{$1^{--}$~state}
The wave function for the $1^{--}$~state can be expanded as
\begin{eqnarray}
\chi(q)=&&(\chi_1(q)+\chi_2(q)\gamma^0+\chi_3(q)(\bbox{\gamma}\cdot{\bf 
q})+ \chi_4(q)(\bbox{\alpha}\cdot{\bf q}))Y_{1m}(\hat{\bf 
q})\nonumber\\
&&{}+(\chi_5(q)\bbox{\gamma}+\chi_6(q)\bbox{\alpha}
+i\chi_7(q)\gamma^5 (\bbox{\alpha}\times{\bf q})+i\chi_8(q)\gamma^5({\bf 
q}\times\bbox{\gamma}))\cdot{\bf 
Y}^{(e)}_{1m}(\hat{\bf q}) .
\end{eqnarray}
The amplitudes are
\begin{eqnarray}
\chi_1(q)=&&(\Delta^{+-}+\Delta^{-+})\frac{1}{2}\Psi_S(q) 
+(\Delta^{+-}-\Delta^{-+})\frac{q}{2E}\Psi_{T_1}(q) 
\nonumber\\
&&{}+(\Delta^{++}+\Delta^{--})\frac{m}{E^2}I_1(q),\\
\chi_2(q)=&&(\Delta^{++}-\Delta^{--})\frac{1}{E}I_1(q),\\
\chi_3(q)=&&(\Delta^{+-}+\Delta^{-+})\frac{m}{2q^2}\Psi_S(q) 
+(\Delta^{+-}-\Delta^{-+})\frac{m}{2Eq}\Psi_{T_1}(q)\nonumber\\
&&{}-(\Delta^{++}+\Delta^{--})\frac{1}{E^2}I_1(q),\\
\chi_4(q)=&&(\Delta^{+-}+\Delta^{-+})\frac{1}{2q}\Psi_{T_1}(q) 
+(\Delta^{+-}-\Delta^{-+})\frac{E}{2q^2}\Psi_{S}(q),\\
\chi_5(q)=&&(\Delta^{+-}+\Delta^{-+})\frac{1}{2}\Psi_V(q)
+(\Delta^{+-}-\Delta^{-+})\frac{E}{2m}\Psi_{T_2}(q),\\
\chi_6(q)=&&(\Delta^{+-}+\Delta^{-+})\frac{1}{2}\Psi_{T_2}(q) 
+(\Delta^{+-}-\Delta^{-+})\frac{m}{2E}\Psi_V(q)\nonumber\\
&&+(\Delta^{++}+\Delta^{--})\frac{q^2}{E^2}I_2(q),\\
\chi_7(q)=&&(\Delta^{++}-\Delta^{--})\frac{1}{E}I_2(q),\\
\chi_8(q)=&&-(\Delta^{+-}+\Delta^{-+})\frac{1}{2m}\Psi_{T_2}(q) 
-(\Delta^{+-}-\Delta^{-+})\frac{1}{2E}\Psi_V(q)\nonumber\\
&&{}+(\Delta^{++}+\Delta^{--})\frac{m}{E^2}I_2(q),
\end{eqnarray}
where
\begin{eqnarray}
I_1(q)=&&mU\Psi_{S}(q)\nonumber\\
&&{}+\frac{1}{2}\int\frac{dp}{(2\pi)^{3}} 
p^2\left(\left(mF_1^S+\frac{mq}{3p} (F_0^S+2F_2^S)\right)\Psi_S(p) 
+q\frac{\sqrt{2}}{3}(F_0^S-F_2^S)\Psi_V(p)\right)\nonumber\\
&&{}+\int\frac{dp}{(2\pi)^{3}} 
p^2\Biggl(\left(2mF_1^V-\frac{mq}{3p}(F_0^V+2F_2^V) +m\frac{q}{2p} 
\left(\frac{q}{p}F_{1}^{Vc}-F_{0}^{Vc}\right)\right)\Psi_S(p) \nonumber\\
&&{}-\left(q\frac{\sqrt{2}}{3}(F_0^V-F_2^V) 
-\frac{(p^{2}-q^{2})}{2\sqrt{2}p}F_{1}^{Vc}\right)\Psi_V(p)\Biggr),\\
I_2(q)=&&-U\Psi_{T_{2}}(q)\nonumber\\
&&{}-\frac{1}{2}\int\frac{dp}{(2\pi)^{3}} 
p^2\left(\frac{\sqrt{2}}{3}(F_0^S-F_2^S)\Psi_{T_1}(p) 
+\left(\frac{p}{q}F_1^S+\frac{1}{3}(2F_0^S+F_2^S)\right)\Psi_{T_2}(p)\right) 
\nonumber\\
&&{}-\int\frac{dp}{(2\pi)^{3}} p^{2}\left( 
\frac{(p^{2}-q^{2})}{2\sqrt{2}pq}F_{1}^{Vc}\Psi_{T_{1}}(p) 
+\left(\frac{p}{q}F_1^V-\frac{1}{3}(F_{0}^{Vc}-F_{2}^{Vc}) 
\right)\Psi_{T_2}(p) \right).
\end{eqnarray}

\subsection*{$0^{++}$~state}
The wave function for the $0^{++}$~state can be expanded as
\begin{equation}
\chi(q)=(\chi_1(q)+\chi_2(q)\gamma^0+\chi_3(q)(\bbox{\gamma}\cdot{\bf 
q})+ \chi_4(q)(\bbox{\alpha}\cdot{\bf q}))Y_{00}(\hat{\bf 
q}).
\end{equation}
The amplitudes are the same as for $1^{--}$ but with $I_1(q)$ given by
\begin{eqnarray}
I_1(q)=&&mU\Psi_{S}(q)+\frac{1}{2}\int\frac{dp}{(2\pi)^{3}} 
p^2\left(mF_0^S+\frac{mq}{p} F_1^S\right)\Psi_S(p) \nonumber\\
&&{}+\int\frac{dp}{(2\pi)^{3}} p^2\left(2mF_0^V-\frac{mq}{p}F_1^V 
+\frac{m}{2}\left(\frac{q}{p}F_{1}^{Vc}-F_{0}^{Vc}\right) 
\right)\Psi_S(p).
\end{eqnarray}

\subsection*{$1^{++}$~state}
The wave function for the $1^{++}$~state can be expanded as
\begin{eqnarray}
\chi(q)=&&(\chi_1(q)\gamma^5+\chi_2(q)\gamma^5\gamma^0+\chi_3(q) 
\gamma^5(\bbox{\gamma}\cdot{\bf q})+ 
\chi_4(q)\gamma^5(\bbox{\alpha}\cdot{\bf q}))Y_{1m}(\hat{\bf 
q})\nonumber\\
&&{}+(\chi_5(q)\gamma^5\bbox{\gamma}+\chi_6(q)\gamma^5
\bbox{\alpha}+i 
\chi_7(q)(\bbox{\alpha}\times{\bf q})+i\chi_8(q)({\bf 
q}\times\bbox{\gamma}))\cdot{\bf 
Y}^{(e)}_{1m}(\hat{\bf q}).
\end{eqnarray}
The amplitudes are
\begin{eqnarray}
\chi_1(q)=&&0,\\
\chi_2(q)=&&(\Delta^{--}-\Delta^{++})\frac{q}{Em}I_1(q),\\
\chi_3(q)=&&(\Delta^{++}+\Delta^{--})\frac{1}{mq}I_1(q),\\
\chi_4(q)=&&(\Delta^{--}-\Delta^{++})\frac{1}{Eq}I_1(q),\\
\chi_5(q)=&&-(\Delta^{+-}-\Delta^{-+})\frac{q}{2E}\Psi_V(q) 
+(\Delta^{+-}+\Delta^{-+})\frac{q}{2m}\Psi_{T}(q)\nonumber\\
&&+(\Delta^{++}+\Delta^{--})\frac{m}{E^2}I_2(q),\\
\chi_6(q)=&&(\Delta^{--}-\Delta^{++})\frac{1}{E}I_2(q),\\
\chi_7(q)=&&-(\Delta^{+-}-\Delta^{-+})\frac{m}{2Eq}\Psi_V(q) 
+(\Delta^{+-}+\Delta^{-+})\frac{1}{2q}\Psi_{T}(q)\nonumber\\
&&{}-(\Delta^{++}+\Delta^{--})\frac{1}{E^2}I_2(q),\\
\chi_8(q)=&&(\Delta^{+-}-\Delta^{-+})\frac{1}{2q}\Psi_{V}(q) 
-(\Delta^{+-}+\Delta^{-+})\frac{E}{2qm}\Psi_T(q),
\end{eqnarray}
where
\begin{eqnarray}
I_1(q)=&&\frac{1}{2}\int\frac{dp}{(2\pi)^{3}} 
p^3\frac{\sqrt{2}}{3}(F_0^S-F_2^S)\Psi_T(p)\nonumber\\
&&{}-\int\frac{dp}{(2\pi)^{3}} 
p^2\left(p\frac{\sqrt{2}}{3}(F_0^V-F_2^V) 
+\frac{(p^{2}-q^{2})}{2\sqrt{2}q}F_{1}^{Vc}\right)\Psi_T(p),\\
I_2(q)=&&qU\Psi_{T}(q)+\frac{1}{2}\int\frac{dp}{(2\pi)^{3}} 
p^2\left(qF_1^S+\frac{p}{3}(2F_0^S+F_2^S)\right) \Psi_T(p)\nonumber\\
&&{}-\int\frac{dp}{(2\pi)^{3}}p^{2}\left(\frac{p}{3}(2F_0^V+F_2^V) 
-\frac{p^{2}}{4q}\left((p^{2}+q^{2})F_{1}^{Vc}-2pq F_{0}^{Vc}\right)
\right)\Psi_T(p).
\end{eqnarray}

\subsection*{$2^{++}$~state} 

The expansion for the $2^{++}$~state is the same as for the 
$1^{--}$~state with the changes $Y_{1m}\to Y_{2m}$, ${\bf 
Y}_{1m}^{(e)}\to{\bf Y}_{2m}^{(e)}$.  The amplitudes are the same but 
with $I_1(q)$ and $I_2(q)$ given by
\begin{eqnarray}
I_1(q)=&&mU\Psi_{S}(q)\nonumber\\
&&{}+\frac{1}{2}\int\frac{dp}{(2\pi)^{3}} 
p^2\left(\left(mF_2^S+\frac{mq}{5p} (2F_1^S+3F_3^S)\right)\Psi_S(p) 
+q\frac{\sqrt{6}}{5}(F_1^S-F_3^S)\Psi_V(p)\right)\nonumber\\
&&{}+\int\frac{dp}{(2\pi)^{3}} p^2\Biggl(\left(2mF_2^V-\frac{mq}{5p} 
(2F_1^V+3F_3^V)+\frac{m}{4p^{2}}\left(( p^{2}+3q^{2})F_{2}^{Vc} -4pq 
F_{1}^{Vc}\right)\right)\Psi_S(p) \nonumber \\
&&{}-\left(q\frac{\sqrt{6}}{5}(F_1^V-F_3^V)-\frac{(p^{2}-q^{2})\sqrt{6}}{4p} 
F_{2}^{Vc}\right)\Psi_V(p)\Biggr),\\
I_2(q)=&&-U\Psi_{T_{2}}(q)\nonumber\\
&&{}-\frac{1}{2}\int\frac{dp}{(2\pi)^{3}} 
p^2\left(\frac{\sqrt{6}}{5}(F_1^S-F_3^S) \Psi_{T_1}(p) 
+\left(\frac{p}{q}F_2^S+ 
\frac{1}{5}(3F_1^S+2F_3^S)\right)\Psi_{T_2}(p)\right) \nonumber\\
&&{}-\int\frac{dp}{(2\pi)^{3}}p^{2} 
\Biggl(\frac{(p^{2}-q^{2})\sqrt{6}}{4pq}F_{2}^{Vc}\Psi_{T_{1}}(p)
\nonumber \\
&&{}+\left(\frac{p}{q}F_2^V+\frac{1}{2pq} \left((p^{2}+q^{2})F_{2}^{Vc} 
-2pq F_{1}^{Vc}\right)\right)\Psi_{T_2}(p)\Biggr).
\end{eqnarray}

\subsection{Lorentz transformed states} 
\label{lorentz}
To calculate the E1 and M1 transitions we need the wave functions for 
some states in a frame where the particle is moving with momentum $P$.

The Salpeter equation in a general frame (\ref{bs-eq}) is found from 
the one in the rest frame as follows.  First this is multiplied by the 
spinor transformation matrices $S(\Lambda)$ and $S^{-1}(\Lambda)$ on 
each side. $\Lambda$ is the Lorentz matrix transforming $(M,{\bf 0})$ 
to the momentum $P$. Then the change of variables 
$q\to\Lambda^{-1}q$ is made.

The Lorentz transformed state is therefore found to be
\begin{equation}
\chi(q,P)=S(\Lambda)\chi(\Lambda^{-1}q,M)S^{-1}(\Lambda),
\end{equation}
where $\chi(q,M)$ is the wave function in the rest frame.  Under the 
change of variables $q\to\Lambda^{-1}q$, then
\begin{eqnarray}
q^0\to&&q_P=\frac{q\cdot P}{M},\\
|{\bf q}|\to&& q_\top=\sqrt{-q^2_\perp},\quad q_\perp=q-\frac{q\cdot 
P}{M^2}P,\\
E\to&&\omega=\sqrt{q_\top^2+m^2},\\
\sin\theta\to&&\sin\theta'= \frac{\sqrt{q_x^2+q_y^2}}{q_\top},\\
\cos\theta\to&&\cos\theta'=\frac{-q^0P_z+q_zP^0}{M q_\top} ,
\end{eqnarray} if the particle moves in the $z$ direction.

For the $0^{-+}$~state the transformed wave function is
\begin{equation}
\chi(q,P)=\left(\chi_A(q)\gamma^5-\chi_B(q)\gamma^5\rlap{/}{q}_\perp 
+\chi_C(q)\gamma^5\frac{\rlap{/}{P}}{M} 
+\chi_D(q)\frac{\gamma^5}{2M}(\rlap{/}{q}_\perp\rlap{/}{P}-
 \rlap{/}{P}\rlap{/}{q}_\perp)\right)Y_{00}(\theta',\varphi).
\end{equation}
The scalar amplitudes are the same as before with the above changes of 
variables, and with the functions $\Psi_i$ depending on $q_\top$.

The Lorentz transformed $1^{--}$ wave function is
\begin{eqnarray}
\chi(q,P)=&&\left(\chi_1+\chi_2\frac{\rlap{/}{P}}{M}- 
\chi_3\rlap{/}{q}_\perp+\chi_4\frac{1}{2M} 
(\rlap{/}{q}_\perp\rlap{/}{P}-\rlap{/}{P}\rlap{/}{q}_\perp)\right) 
Y_{1m}(\theta',\varphi)\nonumber\\
&&{}-\chi_{5}\rlap{/}{Y}+\frac{\chi_{6}}{2M} 
(\rlap{/}{Y}\rlap{/}{P}-\rlap{/}{P}\rlap{/}{Y}) 
+\frac{\chi_{7}}{2}(\rlap{/}{Y}\rlap{/}{q}_\perp- 
\rlap{/}{q}_\perp\rlap{/}{Y}) 
+\chi_{8}\frac{\rlap{/}{P}}{2M}(\rlap{/}{Y} 
\rlap{/}{q}_\perp-\rlap{/}{q}_\perp\rlap{/}{Y}).
\end{eqnarray}
Here
\begin{equation}
Y=\left(\frac{P_z}{M}Y_{z},Y_{x}, 
Y_{y},\frac{P^0}{M}Y_{z}\right)
\end{equation}
with
\begin{equation}
{\bf Y}=(Y_{x}(\theta',\varphi),Y_{y}(\theta',\varphi),Y_{z}(\theta',\varphi)).
\end{equation}

The Lorentz transformed wave functions for $0^{++}$ and $2^{++}$ are 
identical to the one for $1^{--}$ (only the first four terms for 
$0^{++}$).

The wave function for $1^{++}$ is
\begin{eqnarray}
\chi=&&\left(\chi_{1}\gamma^{5}+ \chi_{2}\gamma^{5}\frac{\rlap{/}{P}}{M}
 -\chi_{3}\gamma^{5}
\rlap{/}{q}_\perp +\chi_{4}\frac{\gamma^{5}}{2M}(\rlap{/}{q}_\perp\rlap{/}{P}
-\rlap{/}{P}\rlap{/}{q}_\perp)\right) Y_{1m}(\theta',\varphi) \nonumber \\
&&{}-\chi_{5}\gamma^{5}\rlap{/}{Y} +
\chi_{6}\frac{\gamma^{5}}{2M}(\rlap{/}{Y}\rlap{/}{P}- \rlap{/}{P}\rlap{/}{Y})
+\chi_{7}\frac{\gamma^{5}}{2}(\rlap{/}{Y}\rlap{/}{q}_\perp- 
\rlap{/}{q}_\perp\rlap{/}{Y}) +
\chi_{8}\frac{\gamma^{5}}{2M}\rlap{/}{P}(\rlap{/}{Y}\rlap{/}{q}_\perp- 
\rlap{/}{q}_\perp\rlap{/}{Y}) .
\end{eqnarray}

\section{$\gamma\gamma$ decays} 
\label{ggdecays}
We here present the matrix element for the $\gamma\gamma$ decays 
expressed in the scalar amplitudes defined in 
Appendix~\ref{amplitudes}. We write each scalar amplitude of the wave 
function as
\begin{equation}
\chi_i(q)=\chi_i^{+-}(q)\Delta^{+-}+\chi_i^{-+}(q)\Delta^{-+}+ 
\chi_i^{++}(q)\Delta^{++}+\chi_i^{--}(q)\Delta^{--},
\end{equation}
where the $\Delta^{ij}$'s are defined in (\ref{delta1})-(\ref{delta4}).  
For example in (\ref{D2}) we write
 \begin{equation}
 	\chi_{A}(q)=\chi_{A}^{+-}(q)\Delta^{+-}+\chi_A^{-+}(q)\Delta^{-+}+ 
\chi_A^{++}(q)\Delta^{++}+\chi_A^{--}(q)\Delta^{--},
 \end{equation}
with 
\begin{eqnarray}
	\chi_{A}^{+-} & = & \frac{1}{2}\Psi_{P}(q)+\frac{E}{2m}\Psi_{A}(q),  \\
	\chi_{A}^{-+} & = & \frac{1}{2}\Psi_{P}(q)-\frac{E}{2m}\Psi_{A}(q), \\
	\chi_{A}^{++} & = & \chi_{A}^{--}=0.
\end{eqnarray}
We then take the trace and perform the $q^0$ integration.  This 
gives the following results.  

\subsection*{$0^{-+}$~state}
\begin{eqnarray}
\epsilon_1^\mu\epsilon_2^\nu M_{\mu\nu}=&&-16i\int\frac{d{\bf 
q}}{(2\pi)^{3}} \frac{(\chi_C^{+-}({\bf q}-{\bf k})+\chi_D^{+-}m{\bf 
q})\cdot(\bbox{\epsilon}_1 \times\bbox{\epsilon}_2)}{(M-2E({\bf 
q})-2E({\bf q}-{\bf k}))E({\bf q}-{\bf k})}\nonumber\\
&&{}+16i\int\frac{d{\bf q}}{(2\pi)^{3}} \frac{(\chi_C^{-+}({\bf 
q}-{\bf k})+\chi_D^{-+}m{\bf q})\cdot(\bbox{\epsilon}_1 
\times\bbox{\epsilon}_2)}{(M+2E({\bf q})+2E({\bf q}-{\bf k}))E({\bf 
q}-{\bf k})}\nonumber\\
&&{}+16i\int\frac{d{\bf q}}{(2\pi)^{3}} \frac{1}{(M-2E({\bf q})- 
2E({\bf q}-{\bf k}))(M+2E({\bf q})+2E({\bf q}-{\bf k}))E({\bf q}-{\bf 
k})}\nonumber\\
&&{}\times\biggl[((\chi_B^{--}-\chi_B^{++})E({\bf q}-{\bf k}){\bf q}\nonumber\\
&&{}\quad+ 
(\chi_C^{++}+\chi_C^{--}) ({\bf q}-{\bf k}) 
+(\chi_D^{++}+\chi_D^{--})m{\bf 
q})\cdot(\bbox{\epsilon}_1\times\bbox{\epsilon}_2)\biggr].
\end{eqnarray}

\subsection*{$0^{++}$~state}
\begin{eqnarray}
\epsilon_1^\mu\epsilon_2^\nu M_{\mu\nu}=&&16\int\frac{d{\bf 
q}}{(2\pi)^{3}} \frac{(-m\chi_A^{+-}+{\bf q}\cdot({\bf q}-{\bf 
k})\chi_C^{+-})(\bbox{\epsilon}_1 
\cdot\bbox{\epsilon}_2)-2\chi_C^{+-}(\bbox{\epsilon}_1\cdot{\bf 
q})(\bbox{\epsilon}_2\cdot{\bf q})}{(M- 2E({\bf q})-2E({\bf q}-{\bf 
k}))E({\bf q}-{\bf k})}\nonumber\\
&&{}-16\int\frac{d{\bf q}}{(2\pi)^{3}} \frac{(-m\chi_A^{-+}+{\bf 
q}\cdot({\bf q}-{\bf k})\chi_C^{-+})(\bbox{\epsilon}_1 
\cdot\bbox{\epsilon}_2)- 2\chi_C^{-+}(\bbox{\epsilon}_1\cdot{\bf 
q})(\bbox{\epsilon}_2\cdot{\bf q})}{(M+2E({\bf q})+ 2E({\bf q}-{\bf 
k}))E({\bf q}-{\bf k})}\nonumber\\
&&{}-16\int\frac{d{\bf q}}{(2\pi)^{3}} \frac{1}{(M-2E({\bf q})-2E({\bf 
q}-{\bf k}))(M+2E({\bf q})+2E({\bf q}-{\bf k}))E({\bf q}-{\bf 
k})}\nonumber\\
&&{}\times\biggl[(-m(\chi_A^{++}+\chi_A^{--})+E({\bf q}-{\bf 
k})(\chi_B^{--}-\chi_B^{++})\nonumber\\
&&\quad{} +{\bf q}\cdot({\bf q}-{\bf 
k})(\chi_C^{++}+\chi_C^{--}))(\bbox{\epsilon}_1\cdot\bbox{\epsilon}_2)-2(\chi_C^{++}+\chi_C^{--})(\bbox{\epsilon}_1\cdot{\bf 
q})(\bbox{\epsilon}_2\cdot{\bf q})\biggr].
\end{eqnarray}

\subsection*{$2^{++}$~state}
\begin{eqnarray}
\epsilon_1^\mu\epsilon_2^\nu M_{\mu\nu}=&&16\int\frac{d{\bf 
q}}{(2\pi)^{3}} \frac{1}{(M-2E({\bf q})-2E({\bf q}-{\bf k}))E({\bf 
q}-{\bf k})}\nonumber\\
&&{}\times \biggl[(-m\chi_1^{+-}+{\bf q}\cdot({\bf q}-{\bf 
k})\chi_3^{+-})(\bbox{\epsilon}_1 \cdot\bbox{\epsilon}_2)-2\chi_3^{+-}(\bbox{\epsilon}_1\cdot{\bf 
q})(\bbox{\epsilon}_2\cdot{\bf q})\nonumber\\
&&\quad{}+ \bbox{\chi}_5^{+-}\cdot({\bf 
q}-{\bf k})(\bbox{\epsilon}_1\cdot\bbox{\epsilon}_2)- 
(\bbox{\epsilon}_1\cdot\bbox{\chi}_5^{+-})(\bbox{\epsilon}_2\cdot{\bf 
q})- (\bbox{\epsilon}_1\cdot{\bf 
q})(\bbox{\epsilon}_2\cdot\bbox{\chi}_5^{+-})\biggr]\nonumber\\
&&{}-16\int\frac{d{\bf q}}{(2\pi)^{3}} \frac{1}{(M+2E({\bf q})+2E({\bf 
q}-{\bf k}))E({\bf q}-{\bf k})}\nonumber\\
&&{}\times \biggl[(-m\chi_1^{-+}+{\bf 
q}\cdot({\bf q}-{\bf k})\chi_3^{-+})(\bbox{\epsilon}_1 
\cdot\bbox{\epsilon}_2)-2\chi_3^{-+}(\bbox{\epsilon}_1\cdot{\bf 
q})(\bbox{\epsilon}_2\cdot{\bf q})\nonumber\\
&&\quad{}+ \bbox{\chi}_5^{-+}\cdot({\bf 
q}-{\bf k})(\bbox{\epsilon}_1\cdot\bbox{\epsilon}_2)- 
(\bbox{\epsilon}_1\cdot\bbox{\chi}_5^{-+})(\bbox{\epsilon}_2\cdot{\bf 
q})- (\bbox{\epsilon}_1\cdot{\bf 
q})(\bbox{\epsilon}_2\cdot\bbox{\chi}_5^{-+})\biggr]\nonumber\\
&&{}-16\int\frac{d{\bf q}}{(2\pi)^{3}} \frac{1}{(M-2E({\bf q})-2E({\bf 
q}-{\bf k}))(M+2E({\bf q})+2E({\bf q}-{\bf k}))E({\bf q}-{\bf 
k})}\nonumber\\
&&{}\times\biggl[(-m(\chi_1^{++}+\chi_1^{--})+E({\bf q}-{\bf 
k})(\chi_2^{--}-\chi_2^{++})\nonumber\\
&&\quad{} +{\bf q}\cdot({\bf q}-{\bf 
k})(\chi_3^{++}+\chi_3^{--}))(\bbox{\epsilon}_1\cdot\bbox{\epsilon}_2)
-2(\chi_3^{++}+\chi_3^{--})(\bbox{\epsilon}_1\cdot{\bf 
q})(\bbox{\epsilon}_2\cdot{\bf q})\nonumber\\
&&\quad{}+ 
(\bbox{\chi}_5^{++}+\bbox{\chi}_5^{--})\cdot({\bf q}-{\bf 
k})(\bbox{\epsilon}_1\cdot\bbox{\epsilon}_2)- 
((\bbox{\epsilon}_1\cdot(\bbox{\chi}_5^{++}+\bbox{\chi}_5^{--}))(\bbox{\epsilon}_2\cdot{\bf 
q})\nonumber\\
&&\quad{}+ 
(\bbox{\epsilon}_2\cdot(\bbox{\chi}_5^{++}+\bbox{\chi}_5^{--}))(\bbox{\epsilon}_1\cdot{\bf 
q})) \biggr].
\end{eqnarray}

Here $\bbox{\chi}_{5}(q)=\chi_{5}(q){\bf 
Y}_{2m}^{(e)}(\hat{\bf q})$.

\begin{figure}
	\caption{Feynman diagram for lepton pair decays.}
	\label{leptonfig}
\end{figure}

\begin{figure}
	\caption{Feynman diagrams for two gamma decays.}
	\label{gamgamfig}
\end{figure}

\begin{figure}
	\caption{Feynman diagrams for electromagnetic transitions.}
	\label{m1fig}
\end{figure}

\begin{figure}
	\caption{Amplitudes for the $0^{-+}$ state.  We see that the 
	function $\frac{m}{E}\Psi_{P}-\Psi_{A}$ appearing the negative 
	energy part of the amplitude is much smaller than the function 
	$\frac{m}{E}\Psi_{P}+\Psi_{A}$ in the positive energy part of the 
	amplitude. However, the function $I_{0}/E$ appearing in the mixed energy part is 
	somewhere in between and certainly not negligible compared to the 
	positive energy amplitude.}
	\label{amplitudplot}
\end{figure}

\begin{table}
	\caption{Masses of the charmonium particles.  The four columns in 
	the middle contain the values obtained in the four different fits.  
	F stands for Feynman gauge and C for Coulomb gauge, 10 MeV and 50 
	MeV stands for the minimum value of the experimental error, 
	$\sigma_{i}$ used in the fit.  We see that all four fits give 
	excellent agreement with experiments.  In the last row we give 
	$\chi^{2}$ including all masses using an fictitious error of 10~MeV 
	for all masses in order to make the numbers comparable. 
	We see that the Coulomb gauge gives a slightly better result.  
	Experimental values are taken from Review of Particle Properties 
	data table\protect\cite{datatable}.}
	\begin{tabular}{lddddd}
		
		particle & F 10 MeV & F 50 MeV & C 10 MeV & C 
		50 MeV & exp  \\
		\hline
		$\eta_{c}(1S)$ & 2.97 & 2.97 & 2.97 & 2.97 & 2.98  \\
		
		$\eta_{c}(2S)$[not used in fit] & 3.62 & 3.64 & 3.62 & 3.63 & 3.59  \\
		
		$J/\Psi$ & 3.14 & 3.15 & 3.13 & 3.14 & 3.10  \\
		
		$\Psi(2S)$ & 3.70 & 3.72 & 3.69 & 3.70 & 3.69  \\
		
		$\Psi(2D)$ & 3.76 & 3.77 & 3.76 & 3.77 & 3.77  \\
		
		$\Psi(3S)$ & 4.09 & 4.10 & 4.09 & 4.10 & 4.04  \\
		
		$\Psi(3D)$ & 4.13 & 4.13 & 4.14 & 4.14 & 4.16  \\
		
		$\Psi(4S)$ & 4.39 & 4.39 & 4.40 & 4.39 & 4.42  \\
		
		$\chi_{c0}$ & 3.44 & 3.41 & 3.43 & 3.41 & 3.42  \\
		
		$\chi_{c1}$ & 3.50 & 3.51 & 3.50 & 3.52 & 3.51  \\
		
		$\chi_{c2}$ & 3.49 & 3.50 & 3.49 & 3.51 & 3.56  \\
		
		$h_{c1}$[not used in fit] & 3.49 & 3.50 & 3.49 & 3.51 & 3.53  \\
		\hline
		$\chi^{2}$ & 127 & 153 & 112 & 116 \\
	\end{tabular}
	\protect\label{masstable}
\end{table}
		
\begin{table}
	\caption{Predicted values for the decay widths compared to the 
	experimental values.  The width $J/\Psi\to e^{+}e^{-}$ is used as 
	input for the renormalization.  The row with $\chi^{2}$ values 
	above the lepton decays presents the total $\chi^{2}$ for the two 
	gamma, E1, and M1 transitions, except $\Psi(2S)\to \eta_{c}\gamma$ 
	which would otherwise totally dominate $\chi^{2}$.  The last row 
	is the total $\chi^{2}$ for the lepton decay, but with the decays 
	of the D states excluded, as they depend on higher order 
	corrections not included in our calculations.  Experimental values 
	are taken from Review of Particle Properties data 
	table\protect\cite{datatable} unless otherwise stated.}
	\begin{tabular}{lddddd}
		
		decay & F 10 MeV & F 50 MeV & C 10 MeV & C 
		50 MeV & exp  \\
		\hline
		$\eta_{c}\to \gamma\gamma$ & 6.2 & 6.3 & 6.2 & 6.5 & 7.5${}\pm{}$1.5  \\
		
		$\chi_{c0}\to \gamma\gamma$ & 1.6 & 1.8 & 1.5 & 1.6 & 
		1.7${}\pm{}$1.3\tablenotemark[1] \\
		
		$\chi_{c2}\to \gamma\gamma$ & 0.31 & 0.41 & 0.31 & 0.37 & 
		0.37${}\pm{}$0.17 \\
		
		$J/\Psi\to \eta_{c}\gamma$ & 1.65 & 1.33 & 1.66 & 1.38 & 
		1.14${}\pm{}$0.39 \\
		
		$\Psi(2S)\to \eta_{c}\gamma$ & 9.8 & 10.9 & 10.2 & 12.8 & 
		0.78${}\pm{}$0.24 \\
		
		$\Psi(2S)\to \eta'_{c}\gamma$ & 0.27 & 0.17 & 0.26 & 0.18 & - \\
		
		$\chi_{c0}\to J/\Psi \gamma$ & 130 & 96 & 143 & 110 & 92${}\pm{}$40  \\
		
		$\chi_{c1}\to J/\Psi \gamma$ & 390 & 399 & 426 & 434 & 240${}\pm{}$40  \\
		
		$\chi_{c2}\to J/\Psi \gamma$ & 218 & 195 & 240 & 218 & 267${}\pm{}$33  \\
		
		$\Psi(2S)\to \chi_{c0}\gamma$ & 31 & 47 & 26 & 31 & 26${}\pm{}$4 \\
		
		$\Psi(2S)\to\chi_{c1}\gamma$ & 58 & 49 & 63 & 50 & 24${}\pm{}$4  \\
		
		$\Psi(2S)\to\chi_{c2}\gamma$ & 48 & 47 & 51 & 49 & 22${}\pm{}$4  \\
		\hline
		$\chi^{2}$ & 136 & 127 & 174 & 116 \\
		\hline \hline 
		
		$J/\Psi\to e^{+}e^{-}$ & 5.26 & 5.26 & 5.26 & 5.26 & 5.26${}\pm{}$0.37  \\
		
		$\Psi(2S)\to e^{+}e^{-}$ & 2.81 & 2.45 & 2.92 & 2.68 & 2.14${}\pm{}$0.21  \\
		
		$\Psi(2D)\to e^{+}e^{-}$ & 0.09 & 0.19 & 0.25 & 0.35 & 0.26${}\pm{}$0.04  \\
		
		$\Psi(3S)\to e^{+}e^{-}$ & 1.99 & 1.59 & 2.09 & 1.84 & 0.75${}\pm{}$0.15  \\
		
		$\Psi(3D)\to e^{+}e^{-}$ & 0.14 & 0.29 & 0.05 & 0.07 & 0.77${}\pm{}$0.23  \\
		
		$\Psi(4S)\to e^{+}e^{-}$ & 1.42 & 1.00 & 1.56 & 1.31 & 0.47${}\pm{}$0.10 \\
		\hline
		$\chi^{2}$ & 169 & 62 & 212 & 130 
	\end{tabular}
	\tablenotetext[1]{From Ref.\protect\cite{Cleo}}
	\protect\label{decaytable}
\end{table}

\begin{table}
\caption{Parameter values in the four fits and the value of $\chi^{2}$ 
as defined in Eq.~(\protect\ref{chikvadrat}).  When changing 
$\sigma_{i}$ from 10~MeV to 50~MeV the parameters which change most 
are the quark mass $m$ and the constant $U$. The parameter $\bar \alpha_{s}$ 
naturally changes significantly between the fits in the two different gauges, 
while the other parameters remain almost constant. We see that 
$\chi^{2}$ is slightly smaller in the Coulomb gauge when 
$\sigma_{i}=10$~MeV and slightly smaller in the Feynman gauge when 
$\sigma_{i}=50$~MeV.}
	\begin{tabular}{ldddd}
		
		Parameter & F 10 MeV & F 50 MeV & C 10 MeV & C 50 MeV  \\
		 & $\chi^{2}=110$ & $\chi^{2}={}$10.2 & $\chi^{2}=95$ & 
		$\chi^{2}={}$11.3 \\
		\hline
		$\bar\alpha_{s}$ & 0.282 & 0.288 & 0.348 & 0.369  \\
		
		$m$ & 1.366 & 1.276 & 1.358 & 1.298  \\
		
		$\lambda$ & 0.257 & 0.280 & 0.266 & 0.274  \\
		
		$U$ & $-$0.131 & 0.072 & $-$0.112 & 0.070  
		
	\end{tabular}
	\protect\label{parametertable}
\end{table}

\begin{table}
	\caption{Change of decay amplitudes when only the term with 
	positive energy states is used.  We see that most decay widths 
	would increase, some with several 100\%, when only the  term 
	with positive energy states is used. This shows that it is important 
	to keep all terms in the model.}
	\begin{tabular}{lrrrr}
		
		Decay & F 10 MeV & F 50 MeV & C 10 MeV & C 50 MeV  \\
		\hline
		$J/\Psi\to\eta_{c}\gamma$ & $38\%$ & $66\%$ & $40\%$ & $60\%$  \\
		
		$\Psi(2S)\to\eta_{c}\gamma$ & $-47\%$ & $-67\%$ & $-59\%$ & $-68\%$  \\
		
		$\Psi(2S)\to \eta'_{c}\gamma$ & $117\%$ & $240\%$ & $117\%$ & 
		$212\%$ \\
		
		$\chi_{c0}\to J/\Psi\gamma$ & $185\%$ & $357\%$ & $176\%$ & $307\%$  \\
		
		$\chi_{c1}\to J/\Psi\gamma$ & $15\%$ & $32\%$ & $10\%$ & $21\%$  \\
		
		$\chi_{c2}\to J/\Psi\gamma$ & $58\%$ & $119\%$ & $54\%$ &  $97\%$ \\
		
		$\Psi(2S)\to\chi_{c0}\gamma$ & $65\%$ & $100\%$ & $68\%$ & $101\%$  \\
		
		$\Psi(2S)\to\chi_{c1}\gamma$ & $-17\%$ & $-24\%$ & $-15\%$ & $0\%$  \\
		
		$\Psi(2S)\to\chi_{c2}\gamma$ & $21\%$ & $33\%$ & $18\%$ & $27\%$  
		
	\end{tabular}
		\protect\label{positiveenergytable}
\end{table}

\newpage
\pagestyle{empty}
\addtolength{\oddsidemargin}{-3cm}
\addtolength{\topmargin}{-3cm}

\psfig{file=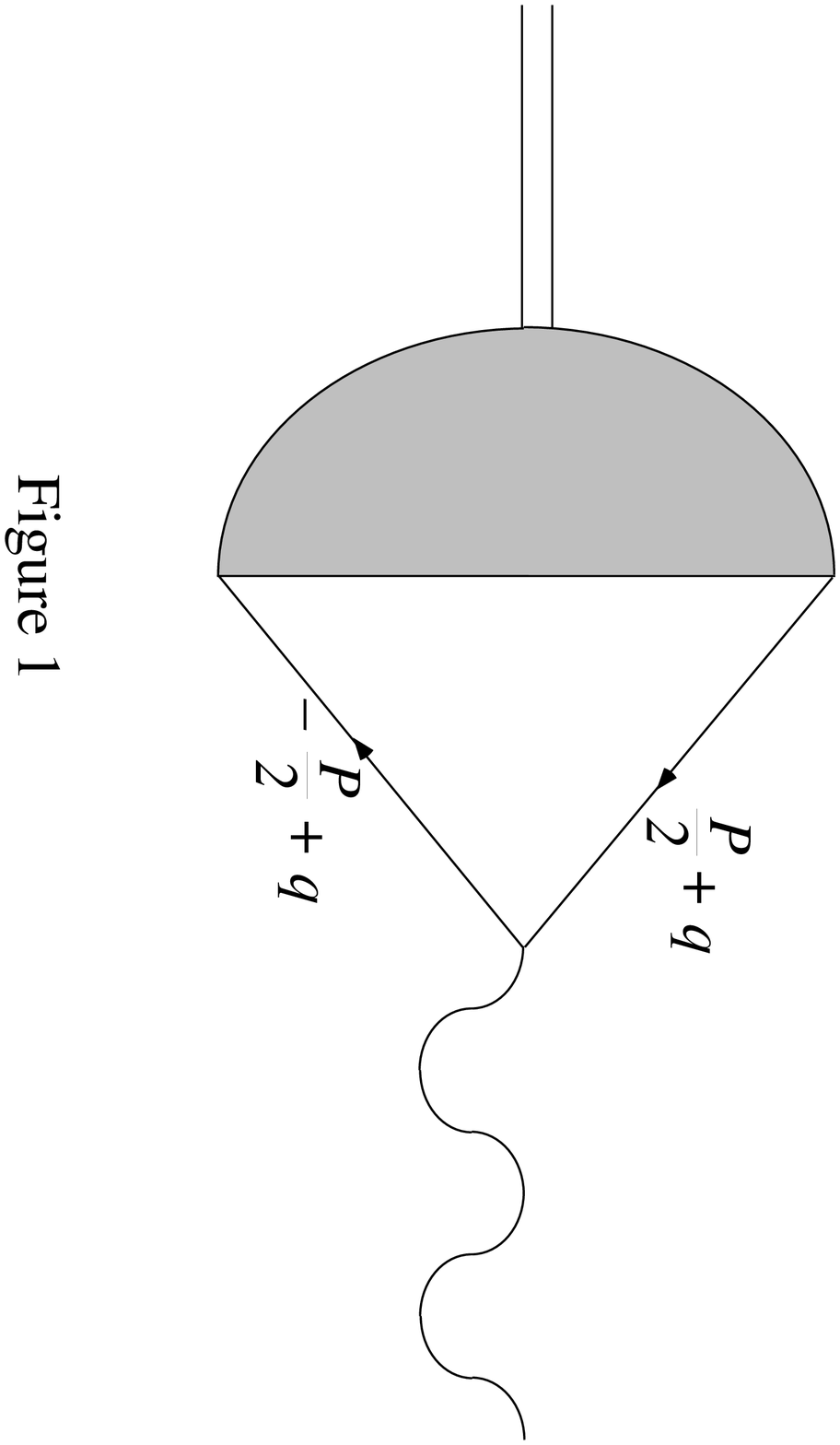}

\newpage

\psfig{file=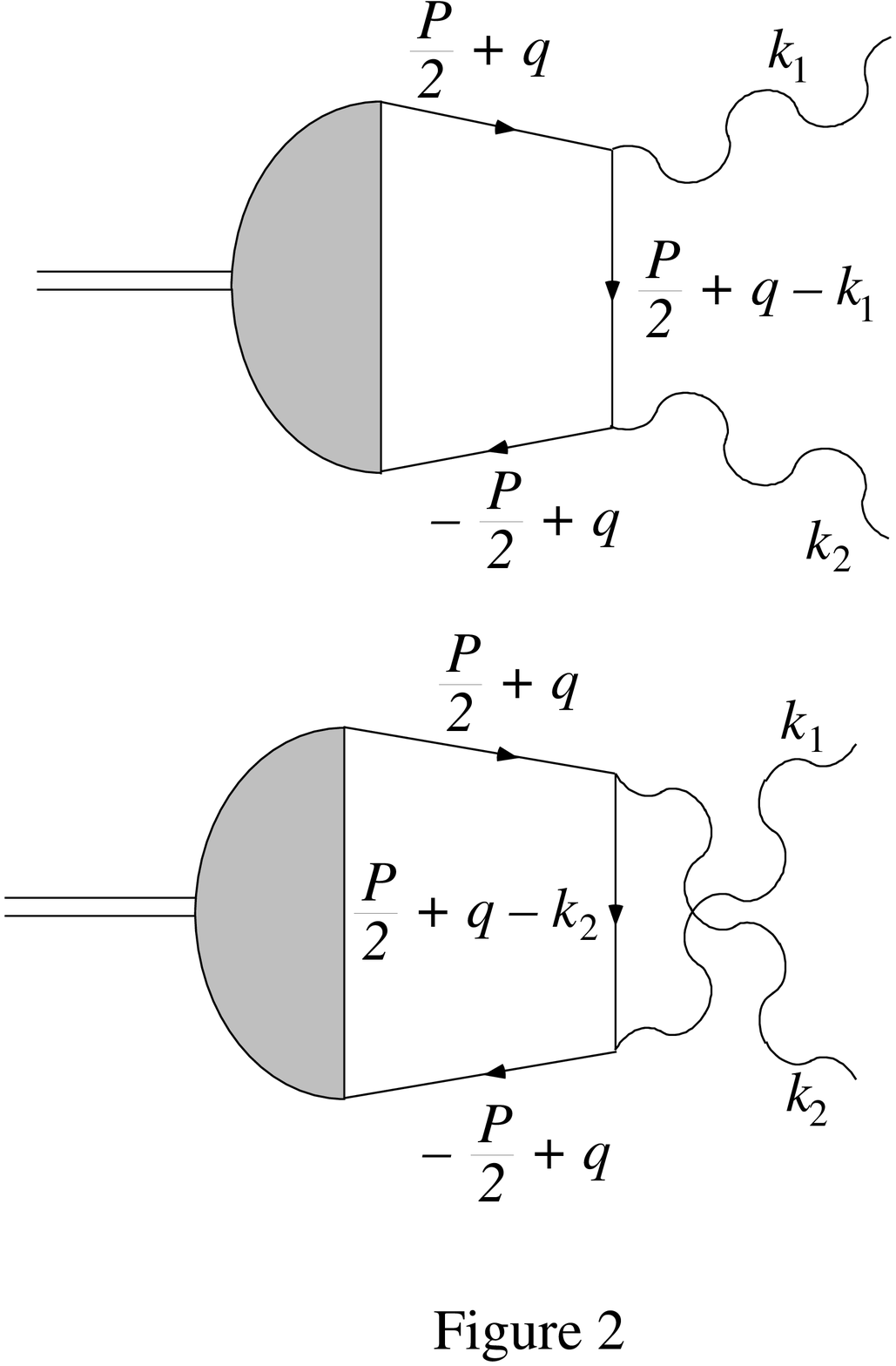}

\newpage

\psfig{file=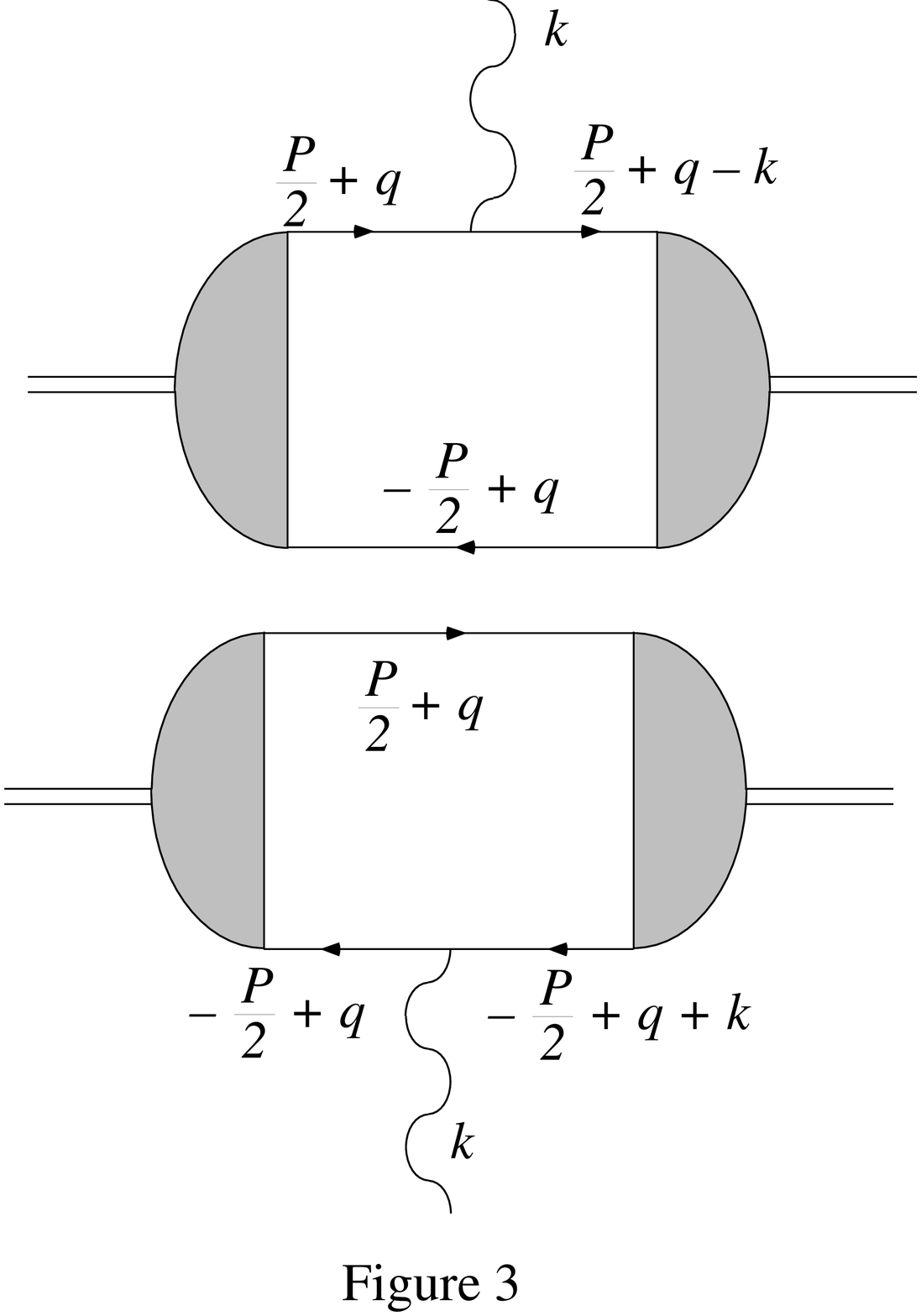}

\newpage

\psfig{file=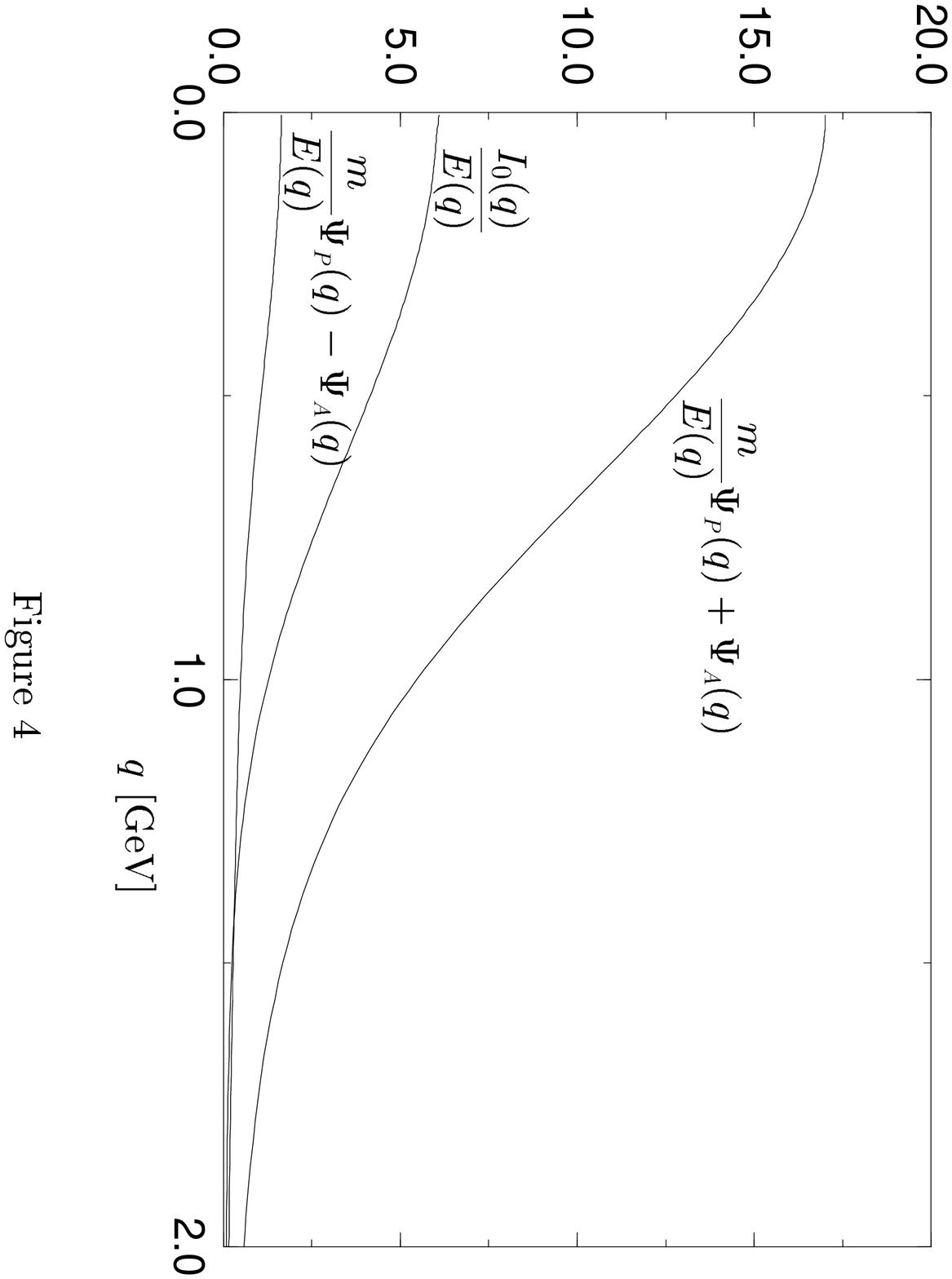}

\end{document}